\documentclass[a4paper,12pt]{article}
\ifx\TeXtures\undefined
\usepackage[active]{srcltx} 
\usepackage[all]{xy}

\usepackage{hyperref}

\hypersetup{ 
colorlinks=true, 
breaklinks=true, 
urlcolor= black, 
linkcolor= black, 
bookmarksopen=true,
pdftitle={}, pdfauthor={}, pdfsubject={}
}

\usepackage[psamsfonts]{amsfonts}%
\usepackage{amsmath}%
\else
\usepackage{amsmath}%
\usepackage{amsfonts}%
\usepackage{times}
\fi
\usepackage{amsthm} 
\usepackage{epsfig}         
\usepackage[utf8]{inputenc}
\usepackage[T1]{fontenc}
\usepackage[english]{babel}
\usepackage{ae,icomma} 
\usepackage{amssymb,amscd,verbatim}
\usepackage{appendix}
\newcommand{\DATUM}{08-02-2024}              
\newcommand{\change}
{{\marginpar{\#}}}        
\newcommand{\comma}{\: ,}              
\newcommand{\period}{\: .}             
\newcommand{\cA}{{\cal A}}
\newcommand{\cB}{{\cal B}}
\newcommand{\cC}{{\cal C}}
\newcommand{\cD}{{\cal D}}
\newcommand{\cE}{{\cal E}}
\newcommand{\cF}{{\cal F}}
\newcommand{\cG}{{\cal G}}

\newcommand{\cI}{{\cal I}}

\newcommand{\cO}{{\cal O}}         

\newcommand{\cR}{{\cal R}}

\newcommand{\cU}{{\cal U}}
\newcommand{\cV}{{\cal V}}
\newcommand{\cW}{{\cal W}}

\newcommand{\cY}{{\cal Y}}

\newcommand{\field}[1]{\mathbb{#1}}
\newcommand{\R}{\field{R}}            
\newcommand{\N}{\field{N}}            
\newcommand{\C}{\field{C}}            

\newcommand{\SSS}{\field{S}}
\newcommand{\ux}{{\underline x}}
\newcommand{\uz}{{\underline z}}

\newcommand{\uxi}{{\underline\xi}}
\newcommand{\rL}{{\rm L}}                 
\newcommand{\rC}{{\rm C}}

\newcommand{\rW}{{\rm W}} 
\newcommand{\rx}{{\rm x}}

\newcommand{\cirS}{\mathop{\bigcirc\kern -.73em {\scriptstyle{\rm S}}}}

\newcommand{\cqfd}{\phantom{blablabla}\hfill\qed\newline} 
\newtheorem{theorem}{Theorem}[section]         
\newtheorem{lemma}[theorem]{Lemma}             
\newtheorem{definition}[theorem]{Definition}   
\newtheorem{remark}[theorem]{Remark}           
\newtheorem{proposition}[theorem]{Proposition} 

\theoremstyle{plain}

\newcommand{\donne}{\mapsto}
\newcommand{\dans}{\longrightarrow}

\newcommand{\Pf}{\vspace*{-2mm}{\bf Proof:}\, }
\newcommand{\Pfof}[1]{{\bf Proof of #1:}\, }

\renewcommand{\theequation}{\thesection.\arabic{equation}}

\newcommand{\ncg}{{[\hskip-.7mm [}}
\newcommand{\ncd}{{]\hskip-.7mm ]}}

\begin{document}

\setcounter{section}{0} 

\title{Limited regularity of a \\
specific electronic reduced density matrix \\
for molecules.}
\author{
{\bf Thierry Jecko}\\
AGM, UMR 8088 du CNRS, site de Saint Martin,\\
2 avenue Adolphe Chauvin,\\
F-95000 Cergy-Pontoise, France. \\
e-mail: jecko@math.cnrs.fr\\
web: http://jecko.perso.math.cnrs.fr/index.html
\\
{\bf Camille Noûs\footnote{The fictitious author Camille Noûs embodies the collegial nature of the present work, as a reminder that science proceeds from disputatio and that the building and dissemination of knowledge are intrinsically selfless, collaborative and open. For details, see:
\[``\ https://www.cogitamus.fr/indexen.html  \ ''\period\]
}}\\
Laboratoire Cogitamus\\
e-mail: camille.nous@cogitamus.fr\\
web: https://www.cogitamus.fr/
}
\date{\DATUM}
\maketitle

\centerline{In memory of {\bf B.T. Sutcliffe} ($\dag$ December 2022.).}

\begin{abstract}
We consider an electronic bound state of the usual, non-relativistic, molecular Hamiltonian with Coulomb interactions, fixed nuclei, and $N$ electrons ($N>1$). Near appropriate electronic collisions, we prove that the $(N-1)$-particle electronic reduced density matrix is not smooth. 
\vspace{2mm}

\end{abstract}

\newpage





\section{Introduction.}
\label{intro}
\setcounter{equation}{0}

The theoretical and practical studies of molecules is known to be an involved task, even for fixed nuclei, since one does not know how to solve the Schrödinger equation for such an electronic system. This explains why an alternative strategy has been developped, namely the Density Functional Theory (DFT) (cf. \cite{e,lise}). To an electronic pure quantum state of the system one attachs several electronic (reduced) density matrices and their regularity properties are important for the DFT. Due to the Coulomb interaction between the particles in the molecule, these regularity properties are not straightforward. In the last two decades, several regularity results were proved on the solution of the Schrödinger equation and on these objects: in particular, 
it has been shown that those electronic density matrices are real analytic in large domains in the configuration space: see \cite{fhhs1,fhhs2,hs1,j1,j2}. This reflects the fact that the potential in the Schrödinger equation is also real analytic on a large domain. However, the optimality of these results is not clear. Intuitively, we do have reasons to think that they are (almost) optimal. A result of this kind has been proved in \cite{fhhs4}. Another one is claimed in \cite{c1} but, as we shall see, it is quite questionable. A more convincing, indirect argument is given in \cite{c2}. In the present paper, we mathematically prove that some particular density matrix is not smooth in some region of the configuration space. 

Let us first recall the mentioned regularity results. We consider a molecule with $N$ moving electrons, with $N>1$, and $L$ fixed nuclei, with $L\geq 1$ (according to Born-Oppenheimer idealization). The $L$ distinct vectors $R_1, \cdots , R_L\in\R^3$ represent the positions of the nuclei. The positions of the electrons are given by $x_1, \cdots , x_N\in\R^3$. The charges of the nuclei are respectively given by the positive $Z_1, \cdots , Z_L$ and the electronic charge is set to $-1$. The Hamiltonian of the electronic system is 
\begin{eqnarray}\label{eq:hamiltonien}
H&:=&\sum _{j=1}^N\Bigl(-\Delta _{x_j}\, -\, \sum _{k=1}^LZ_k|x_j-R_k|^{-1}\Bigr)\, +\, \sum _{1\leq j<j'\leq N}|x_j-x_{j'}|^{-1}\, +\, E_0\comma\hspace{.4cm}\\
\mbox{where}\ E_0&:=&\sum _{1\leq k<k'\leq L}Z_kZ_{k'}|R_k-R_{k'}|^{-1}\nonumber
\end{eqnarray}
and $-\Delta _{x_j}$ stands for the Laplacian in the variable $x_j$. Here we denote by $|\cdot|$ the euclidian norm on $\R^3$. Setting $\Delta :=\sum _{j=1}^N\Delta _{x_j}$, we define the potential $V$ of the system as the multiplication operator satisfying $H=-\Delta +V$. It is well-known that $H$ is a self-adjoint operator on the Sobolev space $\rW ^{2,2}(\R^{3N})$ (cf. Section~\ref{s:basique}). Let us now fix an electronic bound state $\psi\in \rW ^{2,2}(\R^{3N})\setminus\{0\}$ such that, for some real $E$, $H\psi =E\psi$ (there does exist such a state, see Section~\ref{s:basique}). \\
Associated to that bound state $\psi$, we consider the following notions of electronic density. 
Let $k$ be an integer such that $0<k<N$. Let $\rho _k : (\R^3)^k\to\R$ be the almost everywhere defined, $\rL^1(\R^{3k})$-function given by, for $\ux=(x_1; \cdots ; x_k)\in\R^{3k}$, 
\begin{equation}\label{eq:rho-densité}
 \rho _k(\ux)\ =\ \int_{\R^{3(N-k)}}\bigl|\psi (\ux; y)\bigr|^2\, dy\period
\end{equation}
It is called the {\em $k$-particle reduced density}. \\
Define also $\gamma _k : (\R^3)^{2k}\to\C$ as the almost everywhere defined, complex-valued function given by, for $\ux=(x_1; \cdots ; x_k)\in\R^{3k}$ and $\ux'=(x_1'; \cdots ; x_k')\in\R^{3k}$, 
\begin{equation}\label{eq:gamma-densité}
 \gamma _k(\ux; \ux')\ =\ \int_{\R^{3(N-k)}}\overline{\psi (\ux; y)}\, \psi (\ux'; y)\, dy\period
\end{equation}
It is called the {\em $k$-particle reduced density matrix}. \\
Thanks to Kato's important contribution in \cite{k}, we know that the bound state $\psi$ is in fact a continuous function. Therefore, the above densities $\rho _k$ and $\gamma _k$ are actually everywhere defined and continuous, and satisfy $\rho _k(\ux)=\gamma _k(\ux; \ux)$, everywhere. \\
From a physical point of view, the previous objects differ from the true physical ones by some prefactor (see \cite{e,le,lise,lsc}). This will not affect the issue treated here. \\
We need to introduce the following subsets of $\R^{3k}$. Denoting for a positive integer $p$ by $\ncg 1; p\ncd$ the set of the integers $j$ satisfying $1\leq j\leq k$, the closed set 
\begin{equation}\label{eq:coll-elec-k}
 \cC _k\ :=\ \bigl\{\ux=(x_1; \cdots ; x_k)\in\R^{3k}\, ;\, \exists (j; j')\in\ncg 1; k\ncd ^2\, ;\, j\neq j'\ \mbox{and}\ x_j=x_{j'}\bigr\}
\end{equation}
gathers all possible collisions between the first $k$ electrons while the closed set 

\begin{equation}\label{eq:coll-nucl-elec-k}
 \cR _k\ :=\ \bigl\{\ux=(x_1; \cdots ; x_k)\in\R^{3k}\, ;\, \exists j\in\ncg 1; k\ncd \, ,\, \exists \ell\in\ncg 1; L\ncd \, ;\, x_j=R_\ell\bigr\}
\end{equation}
groups together all possible collisions of these $k$ electrons with the nuclei. We set 
\begin{equation}\label{eq:sans-coll-k}
 \cU^{(1)}_k\ :=\ \R^{3k}\setminus \bigl(\cC_k\cup\cR_k\bigr)\comma
\end{equation}
which is an open subset of $\R^{3k}$. \\
The set of all possible collisions between particles is then $\cC_N\cup\cR_N$ and the potential $V$ is real analytic precisely on $\R^{3N}\setminus (\cC_N\cup\cR_N)$. Classical elliptic regularity applied to the equation $H\psi=E\psi$ shows that $\psi$ is also real analytic on $\R^{3N}\setminus (\cC_N\cup\cR_N)$ (cf. \cite{h1}). A better regularity for $\psi$ is not expected (and false in some cases), therefore such a regularity for $\rho _k$ and $\gamma_k$ is not clear. It is however granted on some appropriate region, as stated in Theorem~\ref{th:régu-anal} below. \\
We also need to consider two sets of positions for the first $k$ electrons and introduce the set of all possible collisions between positions of differents sets, namely  
\begin{equation}\label{eq:collisions-extérieures}
 \cC _k^{(2)}\ :=\ \left\{
\begin{array}{l}
 (\ux ; \ux ')\in(\R^{3k})^2\, ;\ \ux=(x_1; \cdots ; x_k)\, ,\ \ux '=(x_1'; \cdots ; x_k')\, ,\, \\
 \\
 \exists (j; j')\in\ncg 1; k\ncd ^2\, ;\  x_j=x_{j'}'
\end{array}
\right\}\period
\end{equation}
We introduce the open subset of $(\R^{3k})^2$ defined by 
\[\cU^{(2)}_k\ :=\ \bigl(\cU^{(1)}_k\times\cU^{(1)}_k\bigr)\setminus\cC _k^{(2)}\period\]
The above mentioned, known regularity results may be summed up in the following way:
\begin{theorem}\label{th:régu-anal}{\rm \cite{fhhs1,fhhs2,hs1,j1,j2}}. \\
For all integer $k$ with $0<k<N$, the $k$-particle reduced density $\rho _k$ is real analytic on $\cU^{(1)}_k$ and the $k$-particle reduced density matrix $\gamma _k$ is real analytic on $\cU^{(2)}_k=(\cU^{(1)}_k\times\cU^{(1)}_k)\setminus\cC _k^{(2)}$. 
\end{theorem}
A natural question is the following: is the domain of real analyticity of $\rho _k$ (resp. $\gamma _k$) larger?
Due to results in \cite{k,fhhs3,fhhs4}, one can show, in the atomic case (i.e. for $L=1$), that $\rho _1$ is not smooth near the nuclear position (see Section~\ref{s:basique}). \\
What about $\gamma _1$? Such a question is considered in the papers \cite{c1,c2}. In \cite{c2}, it is shown that, for two-electron systems in a $S$-state, $\gamma _1$ has a ``fifth order cusp'' on the diagonal, that is a limited regularity there. In \cite{c1}, such a ``fifth order cusp'' on the diagonal is claimed and derived from a special decomposition of the considered bound state near some collisions, that was obtained in \cite{fhhs5} (see also Theorem~\ref{th:décomp-anal} in Section~\ref{ss:décomp-état}). This argument is questionable since the mentioned decomposition is used in \cite{c1} outside the domain of its validity, that was etablished in \cite{fhhs5}. \\
An indirect but interesting approach to show a limited regularity for $\gamma _1$ is introduced in \cite{so1,hs2}. 
The idea is that the lack of smoothness of $\gamma_1$ should be seen in the asymptotics of the eigenvalues of the (class trace) operator with kernel $\gamma _1$. The relevant asymptotics has been obtained in \cite{so1} and it was postulated in Remark 1.2(7) in \cite{hs2} that the leading term is nonzero. This would imply the lack of smoothness of $\gamma_1$ on the diagonal. 
It seems that the non-vanishing of this leading term has not being proved yet. We point out that the same strategy could also work to get the regularity of the one-particle kinetic energy density operator (see \cite{so2}).\\
In the present paper, we take advantage of the above special decomposition to prove that the $(N-1)$-particle electronic reduced density matrix $\gamma _{N-1}$ has a limited regularity at a point on the diagonal of $\cU^{(1)}_{N-1}$. Precisely, we shall show the 
\begin{theorem}\label{th:non-lisse}
Consider  $\hat{\ux}=(\hat{x}_1; \cdots ; \hat{x}_{N-1})\in\cU^{(1)}_{N-1}$  {\rm (}see \eqref{eq:sans-coll-k}{\rm )}. Then the $(N-1)$-particle reduced density matrix $\gamma _{N-1}$ is not smooth near $(\hat{\ux}; \hat{\ux})$. 
\end{theorem}

\begin{remark}\label{r:irrégu-gamma}
We think that Theorem~\ref{th:non-lisse} is valid on a larger region than the diagonal of the set $\cU^{(1)}_{N-1}$. Indeed, we expect that 
the $(N-1)$-particle reduced density matrix $\gamma _{N-1}$ is not smooth near a point $(\hat{\ux}; \hat{\ux}')\in (\cU^{(1)}_{N-1}\times\cU^{(1)}_{N-1})\cap\cC _{N-1}^{(2)}$ {\rm (}see \eqref{eq:sans-coll-k} and \eqref{eq:collisions-extérieures}{\rm )}. We refer to Remark~\ref{r:cas-géné-diag} for an intuitive explanation. 
\end{remark}

A key feature in our proof of that theorem is the splitting of $\gamma _{N-1}$, defined in \eqref{eq:gamma-densité} and restricted to a vicinity of some $(\hat{\ux}; \hat{\ux}')\in (\cU^{(1)}_{N-1}\times\cU^{(1)}_{N-1})\cap\cC _{N-1}^{(2)}$, into a appropriate, finite sum of integrals on regions $\cY$ of $\R^3$ such that, on a neighbourhood of $\hat{\ux}$ times $\cY$ and on a neighbourhood of $\hat{\ux}'$ times $\cY$, the bound state $\psi$ admits a special decomposition as in \cite{fhhs5} (see Theorem~\ref{th:décomp-anal}  in Section~\ref{ss:décomp-état}). Such a decomposition is, up to now, only available at a two-particle collision and this restriction actually explains why our result concerns the $(N-1)$-particle reduced density matrix $\gamma _{N-1}$ and no other $\gamma _k$ and why we focus on collisions $(\hat{\ux}; \hat{\ux}')\in (\cU^{(1)}_{N-1}\times\cU^{(1)}_{N-1})\cap\cC _{N-1}^{(2)}$ (cf. Remark~\ref{r:choix-densité}). We call such a two-particle collision a bilateral collision. \\
It turns out that we need a little more information on this state decomposition near a bilateral collision compared to the one provided by \cite{fhhs5}. This information takes place in Proposition~\ref{prop:terme-non-nul}. \\
Note that there is no restriction on the chosen bound state $\psi$ in Theorem~\ref{th:non-lisse}. As the proof below will show, it however may affect the actual regularity of $\gamma _{N-1}$. More precisely, a connection between this regularity and some characteristic of the bound state decomposition from \cite{fhhs5} will be revealed (see Proposition~\ref{prop:régu-limitée} in Section~\ref{s:2-e} and the proof of Theorem~\ref{th:non-lisse} in Section~\ref{s:general}). We point out that the arguments used in our proofs are rather elementary. \\
In the present paper, we did not address the problem of the possible ``fifth'' order cusp of $\gamma_1$, mentioned in \cite{c2}. We refer to \cite{he} for more information on this question. We think that the technics from the paper \cite{acn} could be useful for this problem. 

It is enlightening to have a look at Theorem~\ref{th:non-lisse} in the simpliest case, namely for $N=2$ and $L=1$. 
In that case, the Hamiltonian $H$ in \eqref{eq:hamiltonien} reduces to the self-adjoint operator acting on $\rW ^{2,2}(\R^{6})\subset\rL ^{2}(\R^{6})$ as 
\begin{equation}\label{eq:hamiltonien-atomique}
 H\ =\ \bigl(-\Delta _x\bigr)\, +\, \bigl(-\Delta _y\bigr)\, +\, \frac{1}{|x\, -\, y|}\, -\, \frac{Z}{|x|}\, -\, \frac{Z}{|y|}\comma
\end{equation}
if the nucleus sits at $0$ in $\R^3$. We have $\cU^{(1)}_1=\R^3\setminus\{0\}$ and 
\[\cU^{(2)}_1\ =\ \bigl(\R^3\setminus\{0\}\bigr)^2\setminus D\comma\]
where $D=\{(x; x')\in(\R^{3})^2; x=x'\}$ is the diagonal of $(\R^{3})^2$. 
For any point $\hat{x}\in\R^3\setminus\{0\}$, Theorem~\ref{th:non-lisse} shows that $\gamma _1$ is not smooth near $(\hat{x}; \hat{x})$. 
However Theorem~\ref{th:régu-anal} tells us that the map $\R^3\ni x\donne\gamma _1(x; x)=\rho _1(x)$ is smooth at such a point $(\hat{x}; \hat{x})$:  $\gamma _1$ is ``smooth along the diagonal''. This behaviour can be easily read off from the already mentioned representation of $\gamma_1$, that crucially relies on the state decomposition at bilateral collisions established in \cite{fhhs5}. From this representation one immediately recovers the smoothness of $\rho _1$ and guesses the non-smoothness of $\gamma _1$ (cf. Remark~\ref{r:régu-rho}).\\
If we consider two electrons but several nuclei at the positions $R_1,\cdots , R_L$ with positive charges $Z_1,\cdots , Z_L$ then the Hamiltonian is given by \eqref{eq:hamiltonien-atomique} with 
\[-\, \frac{Z}{|x|}\, -\, \frac{Z}{|y|}\hspace{.4cm}\mbox{replaced by}\hspace{.4cm}-\, \sum_{\ell=1}^L\, \frac{Z_\ell}{|x\, -\, R_\ell|}\, -\, \sum_{\ell=1}^L\, \frac{Z_\ell}{|y\, -\, R_\ell|}\]
(see \eqref{eq:hamiltonien-2-e}). In this two-electon case, studied in Section~\ref{s:2-e}, $\cU^{(1)}_1=\R^3\setminus\{R_1;\, \cdots ;\, R_L\}$ and 
\[\cU^{(2)}_1\ =\ \bigl(\R^3\setminus\{R_1;\, \cdots ;\, R_L\}\bigr)^2\setminus D\period\]
In particular, the diagonal of $\cU^{(1)}_1$ is precisely $D\cap (\cU^{(1)}_1\times\cU^{(1)}_1)=\cC _1^{(2)}\cap (\cU^{(1)}_1\times\cU^{(1)}_1)$. Theorem~\ref{th:non-lisse} tells us that $\gamma_1$ is not smooth near a point $(\hat{x}; \hat{x})$ on the diagonal $D$, where $\hat{x}\in\R^3\setminus\{R_1;\, \cdots ;\, R_L\}$. \\
We point out that our proof of Theorem~\ref{th:non-lisse} in the general case essentially reduces to the one in the two-electron case. In the latter case, our proof essentially gives more details on the nonsmoothness of $\gamma_1$, this supplementary information being 
a quite precise upper bound on the wave front set (see \cite{h2} for a definition) of $\gamma_1$ above a vicinity of a point $(\hat{x}; \hat{x})$ with $\hat{x}\in\R^3\setminus\{R_1;\, \cdots ;\, R_L\}$. We refer to Proposition~\ref{prop:front-onde} for a precise statement. We further observe, still in the two-electron case, that $\gamma_1$ is the kernel of a pseudo-differential operator on $\R^3\setminus\{R_1;\, \cdots ;\, R_L\}$, the symbol of which belongs to a nice, well-known class of smooth symbols (cf. \cite{h3}): see Proposition~\ref{prop:symbole}. We explain in Section~\ref{s:general} how these two properties extend to the general case. 

The paper is organized as follows: In Section~\ref{s:basique}, we introduce some notation and recall well-known facts on electronic bound states. In Section~\ref{s:bilateral}, we focus on two-particle collisions. We recall in Subsection~\ref{ss:décomp-état} the state decompositions obtained in \cite{fhhs5} and provide the previously mentioned complement on them. In Subsection~\ref{ss:décomp-matrice}, we split $\gamma _{N-1}$ into an appropriate, finite sum of localized integrals and extract from it a smooth part. In Section~\ref{s:2-e}, we focus on the two-electron case. We prove there Theorem~\ref{th:non-lisse} in this case. We also study the wave front set of $\gamma _1$ and the nature of the pseudodifferential operator with $\gamma _1$ as kernel. In Section~\ref{s:general}, we perform the same analysis in the general case and, in particular, prove the main result, namely Theorem~\ref{th:non-lisse}. Technical details are gathered in an Appendix at the end of the paper. \\

{\bf Acknowledgments:} The authors warmly thank L. Garrigue, A. Mizrahi, and R.G. Woolley for fruitful discussions and advice, and also the referee for precise corrections and constructive remarks.

\section{Notation and well-known facts.}
\label{s:basique}
\setcounter{equation}{0}

We start with a general notation. We denote by $\R$ the field of real numbers and by $\C$ the field of complex numbers.  \\
Let $p$ be a positive integer. Recall that, for $u\in\R^p$, we write $|u|$ for the euclidian norm of $u$. Given such a vector $u\in\R^p$ and a nonnegative real number $r$, we denote by $B(u; r[$ (resp. $B(u; r]$) the open (resp. closed) ball of radius $r$ and centre $u$, for the euclidian norm $|\cdot|$ in $\R^p$. \\
In the one dimensional case, we use the following convention for (possibly empty) intervals: for $(a; b)\in\R^2$, let $[a; b]=\{t\in\R; a\leq t\leq b\}$, $[a; b[=\{t\in\R; a\leq t<b\}$, $]a; b]=\{t\in\R; a<t\leq b\}$, and $]a; b[=\{t\in\R; a<t<b\}$.\\
We denote by $\N$ the set of nonnegative integers and set $\N^\ast=\N\setminus\{0\}$. 
If $p\leq q$ are nonnegative integers, we set $\ncg p; q\ncd :=[p; q]\cap\N$, $\ncg p; q\ncg =[p; q[\cap\N$, $\ncd p; q\ncg =]p; q[\cap\N$, and $\ncd p; q\ncd :=]p; q]\cap\N$.\\
Given an open subset $O$ of $\R^p$ and $n\in\N$, we denote by $W^{n,2}(O)$ the standard Sobolev space of those $\rL^2$-functions on $O$ such that, for $n'\in \ncg 0; n\ncd$, their distributional partial derivatives of order $n'$ belong to $\rL^2(O)$. In particular, $W^{0,2}(O)=\rL^2(O)$. Without reference to $O$, we denote by $\|\cdot\|$ (resp. $\langle\cdot , \cdot\rangle$) the $\rL^2$-norm (resp. the right linear scalar product) on $\rL^2(O)$. \\
On $\R^p$, we use a standard notation for partial derivatives. For $j\in\ncg1; p\ncd$, we denote by $\partial _j$ or $\partial _{\rx_j}$ the $j$'th first partial derivative operator. For $\alpha\in\N^p$ and $\rx\in\R^d$, we set $D_\rx^\alpha :=(-i\partial _\rx)^\alpha :=(-i\partial _{\rx_1})^{\alpha _1}\cdots(-i\partial _{\rx_p})^{\alpha _p}$, $D_\rx=-i\nabla_\rx$, $\rx^\alpha :=\rx_1^{\alpha_1}\cdots \rx_p^{\alpha_p}$, $|\alpha |:=\alpha_1+\cdots +\alpha _p$, $\alpha !:=(\alpha_1!)\cdots (\alpha_p !)$, $|\rx|^2=\rx_1^2+\cdots +\rx_p^2$, and $\langle \rx\rangle :=(1+|\rx|^2)^{1/2}$. Given $(\alpha ; \beta)\in(\N^p)^2$, we write $\alpha\leq\beta$ if, for all $j\in\ncg 1; p\ncd$, $\alpha _j\leq\beta_j$. In that case, we define the multiindex $\beta -\alpha:=(\beta_j-\alpha _j)_{j\in\ncg 1; p\ncd}\in \N^p$. \\
We choose the same notation for the length $|\alpha|$ of a multiindex $\alpha\in\N^p$ and for the euclidian norm $|\rx|$ of a vector $\rx\in\R^p$ but the context should avoid any confusion. \\
For $k\in\N\cup\{\infty\}$, we denote by $\rC^k(O)$ the vector space of functions from $O$ to $\C$ which have continuous derivatives up to order $k$ and by $\rC_c^k(O)$ the intersection of $\rC^k(O)$ with the set of function with compact support in $O$. 
If a function $f$ satisfies $f\in\rC^k(O)$ with $k\in\N\cup\{\infty\}$, we often write for this that the function $f$ belongs to the class $\rC^k$ on $O$. In the case $k=\infty$, we also write that $f$ is smooth on $O$ if $f\in\rC^\infty(O)$. \\
Recall that real analytic functions on $O$ are smooth on $O$.
We refer to \cite{h4} for details on the analyticity w.r.t. several variables. \\
In the appendix, we need polynomials. We denote by $\R[X]$ the vector space of the polynomials in one variable with real coefficients. 

Thanks to Hardy's inequality 
\begin{equation}\label{eq:hardy}
\exists\, c>0\, ;\ \forall\, f\in \rW^{1,2}(\R^3)\comma \ \int _{\R^3}|t|^{-2}\, |f(t)|^2\, dt\ \leq \ c\int _{\R^3}|\nabla f(t)|^2\, dt\comma 
\end{equation}
one can show that $V$ is $\Delta$-bounded with relative bound $0$. Therefore the Hamiltonian $H$ is self-adjoint on the domain of the Laplacian $\Delta $, namely $\rW ^{2,2}(\R^{3N})$ (see Kato's theorem in \cite{rs2}, p. 166-167). In particular, for any function $\varphi\in\rW ^{2,2}(\R^{3N})$, each term in the expression of $H\varphi$, that is derived from \eqref{eq:hamiltonien}, makes sense as a $\rL^2$ function on $\R^{3N}$. \\
We point out (cf.\ \cite{si,z}) that a bound state $\psi$ exists at least for appropriate $E\leq E_0$ (cf. \cite{fh}) and for 
\[N\ <\  1\, +\, 2\sum _{k=1}^LZ_k\period\]
A priori, such a bound state $\psi$ just belongs to $\rW ^{2,2}(\R^{3N})$, a space that contains non-continuous functions. But, as shown in $\cite{k}$, $\psi$ is actually continuous. Since the integrand in \eqref{eq:rho-densité} (resp. in \eqref{eq:gamma-densité}) is integrable and continuous, a standard result on the continuity of integrals depending on parameters shows that $\rho _k$ (resp. $\gamma _k$) is everywhere defined and continuous. \\
Kato's paper \cite{k} shows also that $\psi$ has some Hölder continuity (roughly speaking, $\psi$ is almost differentiable). Furthermore, it turns out that singularities of the first derivatives of $\psi$ and also the non-smoothness of $\rho _1$ are encoded in the so called ``cusp condition'' involving some averaged density (see \cite{k,fhhs3}). In \cite{fhhs4}, it is shown, in the atomic case (i.e. for $L=1$), that this averaged density is positive, implying through the cusp condition that $\rho _1$ is not smooth near the nuclear position and that $\psi$ is not differentiable at such place. \\
This low regularity of $\psi$ is due to the collisions of the particles taking place on $\cC_N\cup\cR_N$ (see \eqref{eq:coll-elec-k} and \eqref{eq:coll-nucl-elec-k}). On the complement, the potential $V$ is real analytic and classical elliptic regularity applied to the equation $H\psi=E\psi$ there shows that $\psi$ is also real analytic on $\R^{3N}\setminus (\cC_N\cup\cR_N)$ (cf. \cite{h1}). \\
Another important consequence of the ellipticity of the Hamiltonian $H$ is the following. Let $\delta>0$ and take $\cV_\delta$ a neighbourhood of $\cC_N\cup\cR_N$ in $\R^{3N}$ such that, for $\ux\in\R^{3N}\setminus\cV_\delta$, the distance from $\ux$ to the collisions set $\cC_N\cup\cR_N$ is bigger than $\delta$. On $\R^{3N}\setminus\cV_\delta$, the equation $(H-E)\varphi$, for $\varphi\in\rW ^{2,2}(\R^{3N}\setminus\cV_\delta)$ is an elliptic differential equation with analytic coefficients (see \cite{h3} for a precise definition). Furthermore, the potential $V$ and its derivatives to all orders are bounded on $\R^{3N}\setminus\cV_\delta$ (with bounds depending on $\delta$). 
Starting from $\psi\in\rW ^{2,2}(\R^{3N}\setminus\cV_\delta)$, $(E-V)\psi$ also belongs to $\rW ^{2,2}(\R^{3N}\setminus\cV_\delta)$, thanks to mentioned properties of $V$ on $\R^{3N}\setminus\cV_\delta$. From $(H-E)\psi=0$ on $\R^{3N}\setminus\cV_\delta$, we get $\Delta\psi\in\rW ^{2,2}(\R^{3N}\setminus\cV_\delta)$. This implies that $\psi\in\rW ^{4,2}(\R^{3N}\setminus\cV_\delta)$ (by the lemma on page 52 in \cite{rs2}). Thanks to the nice properties of $V$ on $\R^{3N}\setminus\cV_\delta$ again, $(E-V)\psi$ belongs to $\rW ^{4,2}(\R^{3N}\setminus\cV_\delta)$. This shows $\Delta\psi\in\rW ^{4,2}(\R^{3N}\setminus\cV_\delta)$ and, by ellipticity, $\psi\in\rW ^{6,2}(\R^{3N}\setminus\cV_\delta)$. By induction, we get that $\psi\in\rW ^{2n,2}(\R^{3N}\setminus\cV_\delta)$, for all $n\in\N$. This means, in particular, that any partial derivative $D_\ux^\alpha\psi$ (with $\alpha\in\N^{3N}$) is not only real analytic on $\R^{3N}\setminus\cV_\delta$ but also belongs to $\rL^2(\R^{3N}\setminus\cV_\delta)$. \\
Finally, we further point out that, thanks to Theorem XIII.57, p. 226 in \cite{rs4}, the bound state $\psi$ cannot vanish on a non-empty open subset of $\R^{3N}$. \\
We group those facts together into 
\begin{proposition}\label{prop:faits-état}
Recall that $\cC_N$ {\rm (}resp. $\cR_N${\rm )} is defined in \eqref{eq:coll-elec-k} {\rm (}resp. \eqref{eq:coll-nucl-elec-k}{\rm )}. 
The bound state $\psi$ is a continuous function that also belongs to the Sobolev space $\rW ^{2,2}(\R^{3N})$.
On the open set $\R^{3N}\setminus (\cC_N\cup\cR_N)$, $\psi$ is a real analytic function. Take a subset $\cE$ of $\R^{3N}\setminus (\cC_N\cup\cR_N)$ such that its distance to the collisions set $\cC_N\cup\cR_N$ is positive. Then any partial derivative of $\psi$ belongs to $\rL^2(\cE)$. For any non-empty open set $\cO$ of $\R^{3N}$, the bound state $\psi$ does not vanish identically on $\cO$. 
\end{proposition}
%

\section{Structure of the bound state near bilateral collisions.}
\label{s:bilateral}
\setcounter{equation}{0}

In this section, we recall the decompostion of the considered bound state near a bilateral collision, as obtained in \cite{fhhs5}, and derive its regularity there. We then used such decompositions to split the $(N-1)$-particle reduced density matrix $\gamma _{N-1}$ into a sum of localized integrals. Some of them turn out to be smooth.

\subsection{Decompositions of the bound state.}
\label{ss:décomp-état}

Here we recall the results obtained in Theorem 1.4 in \cite{fhhs5} on the analytic structure of a bound state near a bilateral collision. We add some information on this structure. This allows us to give the regularity of the bound state there. 

First of all, we refer to \cite{h4} for basic notions on (real) analytic functions of several variables. 
With our notation, we rephrase the application of Theorem 1.4 in \cite{fhhs5} on our bound state $\psi$ in the following way: 

\begin{theorem}\label{th:décomp-anal}{\rm \cite{fhhs5}}.\\
The sets $\cC_N$ and $\cR_N$ being defined in \eqref{eq:coll-elec-k} and \eqref{eq:coll-nucl-elec-k}, respectively, we have the following two statements. 
\begin{enumerate}
 \item Consider a point $\hat{\uz}=(\hat{z}_1;\, \cdots\, ;\, \hat{z}_N)\in\R^{3N}\setminus\cC_N$ such that there exists a unique $(j; k)\in\ncg 1;\, N\ncd\times\ncg 1;\, L\ncd$ such that $\hat{z}_j=R_k$. Then there exists a neighbourhood $\Omega$ of $\hat{\uz}$ in $\R^{3N}$ and two real analytic functions $\varphi _1$ and $\varphi _2$ on $\Omega$ such that 
\begin{equation}\label{eq:décomp-noyau}
 \forall \uz\in\Omega\comma\hspace{.4cm}\psi (\uz)\ =\ \varphi _1(\uz)\, +\, |z_j\, -\, R_k|\; \varphi _2(\uz)\period
\end{equation}
 \item Consider a point $\hat{\uz}=(\hat{z}_1;\, \cdots\, ;\, \hat{z}_N)\in\R^{3N}\setminus\cR_N$ such that there exists a unique $(j; k)\in\ncg 1;\, N\ncd^2$ such that $\hat{z}_j=\hat{z}_k$ and $j\neq k$. Then there exists a neighbourhood $\Omega$ of $\hat{\uz}$ in $\R^{3N}$ and two real analytic functions $\varphi _1$ and $\varphi _2$ on $\Omega$ such that 
\begin{equation}\label{eq:décomp-électrons}
 \forall \uz\in\Omega\comma\hspace{.4cm}\psi (\uz)\ =\ \varphi _1(\uz)\, +\, |z_j\, -\, z_k|\; \varphi _2(\uz)\period
\end{equation}
\end{enumerate}
In both cases, we wrote $\uz=(z_1;\, \cdots\, ;\, z_N)\in\R^{3N}$. 
\end{theorem}
In this Theorem~\ref{th:décomp-anal}, we observe that, in both cases, a two-particle collision, or bilateral collision, takes place at $\hat{\uz}$. In the first case, it is an electron-nucleus collision and, in the second one, an electron-electron collision. In both cases, the real analytic functions $\varphi _1$ and $\varphi _2$ a priori depend on $\psi$ and $\hat{\uz}$. They were obtained in \cite{fhhs5} in a somehow abstract way, that was based on the Kustaanheimo-Stiefel transform. \\
Thanks to these decompositions, one can determine the regularity of the bound state $\psi$ near a bilateral collision, as we shall see now. \\
We need to introduce an appropriate notion of {\em valuation} in both cases in Theorem~\ref{th:décomp-anal}. 
Given a nonzero, real analytic function $\varphi$ in several variables $\uz=(z_1;\, \cdots\, ;\, z_N)\in\R^{3N}$, it may be written, near any point $\hat{\uz}=(\hat{z}_1;\, \cdots\, ;\, \hat{z}_N)$ of its domain of analyticity, as the sum of a power series in the variables $((z_1-\hat{z}_1);\, \cdots\, ;\, (z_N-\hat{z}_N))$. For $j\in\ncg 1;\, N\ncd$, this sum may be rearranged in the following form 
\[\varphi (\uz )\ =\ \sum_{\alpha _j\in\N^3}\, \varphi _{\alpha _j}\bigl((z_k)_{k\neq j}\bigr)\; (z_j\, -\, \hat{z}_j)^{\alpha _j}\comma\]
for sums $\varphi _{\alpha _j}$ of appropriate power series in the variables $z_k$ with $k\neq j$. Since the function $\varphi$ is nonzero, so is at least one function $\varphi _{\alpha _j}$. This means that the set $\{|\alpha|;\, \alpha\in\N^3,\, \varphi _\alpha\neq 0\}$ is a non empty subset of $\N$. By definition, the valuation of $\varphi$ in the variable $z_j$ at $\hat{\uz}$ is the minimum of this set. When $\varphi$ is identically zero, we decide to set its valuation in the variable $z_j$ at $\hat{\uz}$ to $-\infty$. 

\begin{definition}\label{def:valuation}
 For the first decomposition of Theorem~\ref{th:décomp-anal}, we define the relevant valuation of the decomposition \eqref{eq:décomp-noyau} at $\hat{\uz}$ as the valuation of $\varphi _2$ in the variable $z_j$ at $\hat{\uz}$. \\
Consider the second decomposition in Theorem~\ref{th:décomp-anal}. We introduce new variables by setting, on $\Omega$, $z_\ell '=z_\ell$, if $\ell\not\in\{j; k\}$, $z_j'=z_j-z_k$, and $z_k'=z_j+z_k$. Replacing each $z_j$ by $\hat{z}_j$, we similarly define the $\hat{z}'_\ell$ from $\hat{z}_\ell$ and set $\hat{\uz}'=(\hat{z}_1';\, \cdots\, ;\, \hat{z}_N')$. The real analytic function $\varphi _2$ on $\Omega$ may be rewritten as $\uz '\donne\tilde\varphi (\uz ')$, for $\uz '=(z_1';\, \cdots\, ;\, z_N')$ in some neighbourhood of $\hat{\uz}'$ and for some real analytic function $\tilde\varphi$ near $\hat{\uz}'$. In that case, we define the relevant valuation of the decomposition \eqref{eq:décomp-électrons} at $\hat{\uz}$ as the valuation of $\tilde\varphi$ in the variable $z_j'$ at $\hat{\uz}'$. 
\end{definition}
This relevant valuation actually governs the regularity of $\psi$ near a bilateral collision, as shown in the 
\begin{proposition}\label{prop:terme-non-nul}
Consider any case in Theorem~\ref{th:décomp-anal} with a small enough open set $\Omega$. The real analytic function $\varphi _2$ cannot be zero identically. In particular, the relevant valuation $q$ of the considered decomposition is an integer. Furthermore, on $\Omega$, the bound state $\psi$ belongs to the class $\rC^q$ but does not belong to the class $\rC^{q+1}$. 
\end{proposition}
\Pf Let us consider the first case in Theorem~\ref{th:décomp-anal}. We may assume that, on $\Omega$, only the collision $z_j=R_k$ occurs. Thus, on $\Omega$, we have $H\psi=E\psi$ where the potential $V$ is given by 
$Z_k/|z_j-R_k|$ plus a real analytic term. \\
Assume that $\varphi _2$ is zero identically on $\Omega$. Inserting the decomposition of $\psi$ into $H\psi=E\psi$, we see that $\uz\donne\varphi _1(\uz)Z_k/|z_j-R_k|$ is almost everywhere equal to some real analytic function $\varphi _3$ on $\Omega$. Thus, by continuity, $Z_k\varphi _1(\uz)=|z_j-R_k|\varphi _3(\uz)$ everywhere on $\Omega$.\\
We use the elementary
\begin{lemma}\label{lm:régu-valuation}
Let $\cW$ be a bounded neighbourhood of $0$ in $\R^3$ and $\varphi :\cW\dans\C$ be a nonzero real analytic function with valuation $q\in\N$ {\rm (}w.r.t. its $3$-dimensional variable in the above sense{\rm )}. Then the function $N_\varphi : \cW\ni x\donne |x|\, \varphi (x)\in\C$ belongs to the class $\rC^q$ but does not belong to the class $\rC^{q+1}$. Furthermore, any partial derivative of order $q+1$ of $N_\varphi$ is well-defined away from zero and bounded. 
\end{lemma}
\Pf See in the Appendix. $\cqfd$

Suppose that $\varphi _3$ is not zero identically on $\Omega$. Then, there exists some 
\begin{equation}\label{eq:un-z}
 \uz\, =\, \bigl(z_1;\, \cdots ;\, z_{j-1};\,  R_k;\,  z_{j+1};\, \cdots ;\, z_N\bigr)\, \in\, \Omega
\end{equation}
such that the map
\[x\ \donne\ \varphi _3\bigl(z_1;\, \cdots\, ;\, z_{j-1};\, x+R_k;\, z_{j+1};\, \cdots\, ;\, z_N\bigr)\]
is not zero identically near $0$. Applying Lemma~\ref{lm:régu-valuation} to this function, we get that $\uz\donne\varphi _1(\uz)=Z_k^{-1}|z_j-R_k|\varphi _3(\uz)$ is not smooth as a function of $z_j$. This contradicts the real analyticity of $\varphi _1$. Thus, on $\Omega$, the function $\varphi _3$ is identically zero and so is $\varphi _1$ too. This implies that the bound state $\psi$ is zero on some non-empty open set. This is now a contradiction with Proposition~\ref{prop:faits-état}. Therefore $\varphi _2$ cannot be zero on $\Omega$. \\
By Lemma~\ref{lm:régu-valuation} and the decomposition \eqref{eq:décomp-noyau} of $\psi$, we immediately see that $\psi$ has the $C^q$ regularity on $\Omega$. Assume now that $\psi$ belongs to the class $\rC^{q+1}$ on $\Omega$. By the definition of the relevant valuation of this decomposition \eqref{eq:décomp-noyau}, we can find some $\uz$ as in \eqref{eq:un-z} such that the function 
\[x\donne \varphi _2\bigl(z_1; \cdots ; z_{j-1}; x+R_k; z_{j+1}\cdots z_N\bigr)\]
is not identically zero near $0$ and its valuation there is $q$. By the decomposition \eqref{eq:décomp-noyau} and the assumption on $\psi$, the function 
\[x\donne |x|\, \varphi _2\bigl(z_1; \cdots ; z_{j-1}; x+R_k; z_{j+1}\cdots z_N\bigr)\]
belongs to the class $\rC^{q+1}$. This contradicts Lemma~\ref{lm:régu-valuation}. \\
Let us consider the second case in Theorem~\ref{th:décomp-anal}. Here also, we may assume that, on $\Omega$, only one collision occurs, say between $z_j$ and $z_k$ (with $k\neq j$ in $\ncg 1; N\ncd$). This means that, on $\Omega$, the potential $V$ is given by $1/|z_j-z_k|$ plus a real analytic term. We can adapt the previous arguments to show the desired result in the second case. $\cqfd$

\begin{remark}\label{r:symmétries}
 Consider the second case in Theorem~\ref{th:décomp-anal} for $N=2$. If we take a fermionic {\rm (}resp. bosonic{\rm )} bound state $\psi$, then we can check with the help of the proof of Proposition~\ref{prop:terme-non-nul} that the functions $\varphi _1$ and $\varphi _2$ are also fermionic {\rm (}resp. bosonic{\rm )}. This implies, in particular, that the relevant valuation is odd {\rm (}resp. even{\rm )}. 
\end{remark}
%

\subsection{Decomposition of the density matrix.}
\label{ss:décomp-matrice}

Making use of the decompositions in Theorem~\ref{th:décomp-anal} at a bilateral collision, we extract here, in a relevant region, a smooth contribution from the $(N-1)$-particle reduced density matrix $\gamma _{N-1}$.

Recall that we introduced the set $\cU^{(1)}_{N-1}$ in \eqref{eq:sans-coll-k} and the set $\cC _{N-1}^{(2)}$ in \eqref{eq:collisions-extérieures}. Let us take $(\hat{\ux}; \hat{\ux}')\in (\cU^{(1)}_{N-1}\times\cU^{(1)}_{N-1})\, \cap\, \cC _{N-1}^{(2)}$. As vectors in $\R^{3(N-1)}$, we write $\hat{\ux}=(\hat{x}_1;\, \cdots;\, \hat{x}_{N-1})$ and $\hat{\ux}'=(\hat{x}_1';\, \cdots;\, \hat{x}_{N-1}')$. Consider the set 
\[\cG\ :=\ \bigl\{(j; j')\in\ncg 1;\, N-1\ncd^2\, ;\ \hat{x}_j\, =\, \hat{x}_{j'}'\bigr\}\period\]
Since $(\hat{\ux}; \hat{\ux}')\in \cC _{N-1}^{(2)}$, $\cG$ is not empty. Since $\hat{\ux}\in\cU^{(1)}_{N-1}$ and $\hat{\ux}'\in\cU^{(1)}_{N-1}$, $\cG$ is the graph of an injective map $c : \cD\dans\ncg 1;\, N-1\ncd$ with the domain of definition 
\[\cD\ :=\ \bigl\{j\in\ncg 1;\, N-1\ncd\, ;\ \exists\, j'\in\ncg 1;\, N-1\ncd\, ;\ (j; j')\in\cG\bigr\}\ \neq\ \emptyset\period\]
\begin{proposition}\label{prop:mod-C-infini}
 Let $(\hat{\ux}; \hat{\ux}')\in (\cU^{(1)}_{N-1}\times\cU^{(1)}_{N-1})\, \cap\, \cC _{N-1}^{(2)}$. Then there exist a neighbourhood $\cV$ of $\hat{\ux}$, a neighbourhood $\cV'$ of $\hat{\ux}'$, a smooth function $s: \cV\times\cV'\dans\C$, and, for $j\in\cD$, a neighbourhood $\cW_j$ of $\hat{x}_j$, a function $\chi _j\in\rC_c^\infty(\R^3; \R^+)$, and two real analytic functions $\varphi _j : \cV\times \cW_j\dans\C$ and $\varphi _{c(j)}' : \cV'\times \cW_j\dans\C$, such that $\chi _j=1$ near $\hat{x}_j=\hat{x}_{c(j)}'$ and the support of $\chi _j$ is included in $\cW_j$, and such that, for all $(\ux; \ux')\in (\cV\times\cV')$, 
 \begin{equation}\label{eq:mod-C-infini}
 \gamma _{N-1}(\ux ;\, \ux ')\ =\ s(\ux ;\, \ux ')\, +\, \sum _{j\in\cD}\, \int_{\R^3}\, |x_j\, -\, y|\, \overline{\varphi_j}(\ux ; y)\, |x_{c(j)}'\, -\, y|\, \varphi_{c(j)}'(\ux '; y)\chi _j(y)\, dy\period 
 \end{equation}
\end{proposition}
\Pf We observe that the points $\hat{x}_j$, for $j\in\ncg 1;\, N-1\ncd$, $\hat{x}_{j'}'$, for $j'\in\ncg 1;\, N-1\ncd\setminus c(\cD)$, and $R_k$, for $k\in\ncg 1;\, L\ncd$, are pairwise distinct. Therefore, we can find a neighbourhood $\cV$ of $\hat{\ux}$, a neighbourhood $\cV'$ of $\hat{\ux}'$, and, on $\R^3$, a partition of unity  
\begin{equation}\label{eq:partition}
 1\ =\ \tau\, +\, \sum _{k=1}^L\, \tau _k\, +\, \sum_{j\in\cD}\, \chi _j\, +\, \sum_{j\not\in\cD}\, \rho_j\, +\, \sum_{j'\not\in c(\cD)}\, \rho_{j'}'\comma
\end{equation}
satisfying the following properties. Let 
\[\cF\ :=\ \{\tau\}\, \cup\, \bigl\{\tau _k\, ;\ k\in\ncg 1;\, L\ncd\bigr\}\, \cup\, \bigl\{\chi _j\, ;\ j\in\cD\bigr\}\, \cup\, \bigl\{\rho _j\, ;\ j\not\in\cD\bigr\}\, \cup\, \bigl\{\rho _{j'}'\, ;\ j'\not\in c(\cD)\bigr\}\period\]
\begin{itemize}
 \item[a).] For any $\eta\in\cF\setminus\{\tau\}$, $\eta\in\rC_c^\infty(\R^3)$, $\tau$ is smooth, and both are nonnegative. 
 \item[b).] For any $(\eta ;\, \theta)\in(\cF\setminus\{\tau\})^2$ with $\eta\neq\theta$, $\eta\, \theta=0$. 
 \item[c).] For $k\in\ncg 1;\, L\ncd$, $\tau _k=1$ near $R_k$. For $(\ux ; \ux ')\in\cV\times\cV'$ and $y$ in the support of $\tau_k$, $(\ux ; y)$ (resp. $(\ux '; y)$) belongs to a neighbourhood $\Omega^{(k)}$ (resp. $(\Omega ')^{(k)}$) of $\hat{\uz }^{(k)}:=(\hat{\ux}; R_k)$ (resp. $\hat{\uz }^{'(k)}:=(\hat{\ux}'; R_k)$), on which the decomposition \eqref{eq:décomp-noyau} with $j=N$ holds true.
 \item[d).] For $j\in\ncg 1;\, N-1\ncd\setminus\cD$, $\rho _j=1$ near $\hat{x}_j$. For $(\ux ; \ux ')\in\cV\times\cV'$ and $y$ in the support of $\rho _j$, $(\ux ; y)$ belongs to a neighbourhood $\Omega_{(j)}$ of $\hat{\uz}:=(\hat{\ux}; \hat{x}_j)$, on which the decomposition \eqref{eq:décomp-électrons} with $k=N$ holds true, and $(\ux '; y)$ belongs to $\R^{3N}\setminus (\cC_N\cup\cR_N)$. 
 \item[e).] For $j'\in\ncg 1;\, N-1\ncd$ with $j'\not\in c(\cD)$, $\rho _{j'}'=1$ near $\hat{x}_{j'}'$. For $(\ux ; \ux ')\in\cV\times\cV'$ and $y$ in the support of $\rho _{j'}'$, $(\ux '; y)$ belongs to a neighbourhood $\Omega_{(j')}'$ of $\hat{\uz}:=(\hat{\ux}'; \hat{x}_{j'}')$, on which the decomposition \eqref{eq:décomp-électrons} with $j=j'$ and $k=N$ holds true, and $(\ux ; y)$ belongs to $\R^{3N}\setminus (\cC_N\cup\cR_N)$. 
 \item[f).] For $j\in\cD$, $\chi _j=1$ near $\hat{x}_j$. For $(\ux ; \ux ')\in\cV\times\cV'$ and $y$ in the support of $\chi _j$, $(\ux ; y)$ belongs to a neighbourhood $\Omega_j$ of $\hat{\uz}:=(\hat{\ux}; \hat{x}_j)$, on which the decomposition \eqref{eq:décomp-électrons} for $(j; N)$ holds true, and $(\ux '; y)$ belongs to a neighbourhood $\Omega_j'$ of $\hat{\uz}':=(\hat{\ux}'; \hat{x}_j)$, on which the decomposition \eqref{eq:décomp-électrons} for $(c(j); N)$ holds true. 
 \item[g).] We have $\cD =\ncg 1; N-1\ncd$ if and only if $c(\cD) =\ncg 1; N-1\ncd$. In that case, the above conditions d) and e) are empty. 
\end{itemize}
Now, we insert \eqref{eq:partition} into the integral of \eqref{eq:gamma-densité} for $k=N-1$. 
On $\cV\times\cV'$, we thus have 
\begin{equation}\label{eq:somme-gamma}
 \gamma _{N-1}(\ux ;\, \ux ')\ =\ \sum _{\eta\in\cF}\, \cI _\eta(\ux ;\, \ux ')\comma\ \mbox{where}\hspace{.4cm}\cI _\eta(\ux ;\, \ux ')\ :=\ \int_{\R^3}\, \overline{\psi}(\ux ;\, y)\; \psi (\ux ' ;\, y)\; \eta (y)\, dy\period
\end{equation}
We separately study the above integrals. We start with $\cI_\tau$. Denoting by $S_\tau$ the support of $\tau$, the distance from $\cV\times S_\tau$ to the collisions set $\cC_N\cup\cR_N$ is positive and so is also the one from $\cV'\times S_\tau$. By Proposition~\ref{prop:faits-état}, the derivatives $\partial _\ux^\alpha\psi$ are continuous, $\rL^2(S_\tau)$-valued functions. Thus, by induction and standard derivation under the integral sign, the partial derivatives of $\cI_\tau$ w.r.t. the variable $(\ux ;\, \ux ')$ all exist and are continuous, yielding the smoothness of $\cI_\tau$. \\
For functions $f$ and $g$ of the variable $(\ux ;\, \ux ')$, we write $f\sim g$ when $f-g$ is smooth. 
We recall that, for $\eta\in(\cF\setminus\{\tau\})$, $\eta$ has a compact support denoted by $S_\eta$. \\
Let $k\in\ncg 1;\, L\ncd$. Using the decompositions mentioned in c), one can use standard derivation under the integral sign to show that $\cI _{\tau _k}$ is smooth. \\
Let $j\in\ncg 1;\, N-1\ncd\setminus\cD$. Making use of the decomposition mentioned in d), there exist real analytic functions $\varphi _1$ and $\varphi _2$ such that, for $(\ux ;\, \ux ')\in\cV\times\cV'$, 
\[\cI _{\rho _j}(\ux ;\, \ux ')\ = \ \int_{\R^3}\, \bigl(\overline{\varphi _1}(\ux ; y)\, +\, \overline{\varphi _2}(\ux ; y)\; |x_j\, -\, y|\bigr)\; \psi (\ux '; y)\; \rho _j(y)\, dy\period\]
Furthermore, $\psi$ is smooth (even real analytic) on $\cV'\times S_{\rho _j}$, by Proposition~\ref{prop:faits-état}. \\
Applying standard derivation under the integral sign, we get, for $(\ux ;\, \ux ')\in\cV\times\cV'$, 
\begin{align*}
\cI _{\rho _j}(\ux ;\, \ux ')\ \sim& \ \int_{\R^3}\, |x_j\, -\, y|\; \overline{\varphi _2}(\ux ; y)\; \psi (\ux '; y)\; \rho _j(y)\, dy\\
\ \sim& \ \int_{\R^3}\, |y'|\; \overline{\varphi _2}\bigl(\ux ; y'+x_j\bigr)\; \psi \bigl(\ux '; y+x_j\bigr)\; \rho _j\bigl(y'+x_j\bigr)\, dy'\comma
\end{align*}
by a change of variables. Again, standard derivation under the integral sign shows that $\cI _{\rho _j}$ is smooth. \\
We treat in the similar way the integrals $\cI _{\rho _{j'}'}$, for $j'\not\in c(\cD)$, by exchanging the rôles of the variables $\ux$ and $\ux'$. \\
We are left with the $\cI _{\chi _j}$, for $j\in\cD$. For such $j$, we use f). There exist four real analytic functions $\varphi _1$, $\varphi _2$, $\varphi _1'$, and $\varphi _2'$, such that, for $(\ux ;\, \ux ')\in\cV\times\cV'$, 
\[\cI _{\chi _j}(\ux ;\, \ux ')\ = \ \int_{\R^3}\, \Bigl(\overline{\varphi _1}(\ux ; y)\, +\, \overline{\varphi _2}(\ux ; y)\; |x_j\, -\, y|\Bigr)\; \Bigl(\varphi _1'(\ux '; y)\, +\, \varphi _2'(\ux '; y)\; |x_{c(j)}'\, -\, y|\Bigr)\; \chi _j(y)\, dy\period\]
Using standard derivation under the integral sign, we obtain 
\begin{align*}
\cI _{\chi _j}(\ux ;\, \ux ')\ \sim& \ \int_{\R^3}\, |x_j\, -\, y|\; \overline{\varphi _2}(\ux ; y)\; \varphi _1' (\ux '; y)\; \chi _j(y)\, dy\\
\ & \ +\, \int_{\R^3}\, |x_{c(j)}'\, -\, y|\; \overline{\varphi _1}(\ux ; y)\; \varphi _2' (\ux '; y)\; \chi _j(y)\, dy\\
\ & \ +\, \int_{\R^3}\, |x_j\, -\, y|\; |x_{c(j)}'\, -\, y|\; \overline{\varphi _2}(\ux ; y)\; \varphi _2' (\ux '; y)\; \chi _j(y)\, dy\period
\end{align*}
Making use of a change of variables as above, we see that the first two integrals on the r.h.s. are smooth, yielding 
\[\cI _{\chi _j}(\ux ;\, \ux ')\ \sim \ \int_{\R^3}\, |x_j\, -\, y|\; |x_{c(j)}'\, -\, y|\; \overline{\varphi _2}(\ux ; y)\; \varphi _2' (\ux '; y)\; \chi _j(y)\, dy\period\]
We have proved \eqref{eq:mod-C-infini}. $\cqfd$

\begin{remark}\label{r:choix-densité} We comment on our proof of Proposition~\ref{prop:mod-C-infini}. \\
In this proof, we crucially use the fact that each factor $\psi $ in some integrals in \eqref{eq:somme-gamma} may be decomposed as in \eqref{eq:décomp-électrons} in Theorem~\ref{th:décomp-anal}. This is justified only at bilateral collisions. If we look at $\gamma_k$ with $k<N-1$ and a fixed $(\ux ; \ux ')$ {\rm (}cf. \eqref{eq:gamma-densité}{\rm )} then, on the domain of integration, there is a $p$-particle collision with $p\geq 3$ when two different, $3$-dimensional $y$-variables meet some $x_j$. Since we do not know how to handle this situation, we restrict ourselves to the case $k=N-1$. \\
Even in that case $k=N-1$, we also get in \eqref{eq:gamma-densité} a $p$-particle collision with $p\geq 3$, if $x_1=x_2$, when one $3$-dimensional $y$-variable meets $x_1$. This explains why we took a point $\hat{\ux}\in\cU^{(1)}_{N-1}$ in Theorem~\ref{th:non-lisse}.\\
We note that the non-smooth decomposition \eqref{eq:décomp-noyau} produces smooth contributions to the density $\gamma_{N-1}$ {\rm (}cf. the contribution of the $\tau _k${\rm )}. \\
When $N>2$ and, for instance, $\hat{x}_1\neq\hat{x}_1'$, we observe that the state decompositions used in f) may be different since they take place near different points $\hat{\uz}$ and $\hat{\uz}'$. This phenomenon is absent if we require the equality $\hat{\ux}=\hat{\ux}'$. Indeed, in that case, $\cD=\ncg 1; N-1\ncd$ and $c$ is the identity map on $\cD$. Therefore, for all $j\in\cD$, the functions $\varphi _2$ and $\varphi _2'$, appearing in f), are equal. \\
For further purpose, we observe that we may require, in the statement of Proposition~\ref{prop:mod-C-infini}, that, for all $j\in\cD$, the functions $\varphi _j$ and $\varphi _{c(j)}'$ are the sum of a power series. If necessary, it suffices to shrink the neighbourhoods $\cV$, $\cV'$, and $\cW_j$, in the proof of Proposition~\ref{prop:mod-C-infini}. 
\end{remark}
%

\section{The two-electron case.}
\label{s:2-e}
\setcounter{equation}{0}

In this section, we focus on the two-electron case. It is already an interesting case but we also shall show below that our result in the general case essentially reduces to it. 

Recall that, for $N=2$, the electronic Hamiltonian is given by 

\begin{equation}\label{eq:hamiltonien-2-e}
 H\ =\ \bigl(-\Delta _x\bigr)\, +\, \bigl(-\Delta _y\bigr)\, +\, \frac{1}{|x\, -\, y|}\, -\, \sum_{k=1}^L\frac{Z_k}{|x\, -\, R_k|}\, -\, \sum_{k=1}^L\frac{Z_k}{|y\, -\, R_k|}\comma
\end{equation}
for positive $Z_1,\, \cdots\, ,\, Z_L$. We have $\cC_1=\emptyset$ thus $\cU^{(1)}_1=\R^3\setminus\{R_1;\, \cdots\, ;\, R_L\}$ (cf. \eqref{eq:sans-coll-k}). The set $\cC_1^{(2)}$ (cf. \eqref{eq:collisions-extérieures}) is precisely the diagonal of $(\R^{3})^2$, that is the set 
\[D\ :=\ \bigl\{(x; x')\in(\R^{3})^2;\, x=x'\bigr\}\comma\hspace{.4cm}\mbox{and}\hspace{.4cm}\cU^{(2)}_1\ =\ \bigl(\cU^{(1)}_1\bigr)^2\setminus D\period\]
We want to consider the situation of Theorem~\ref{th:non-lisse} with $N=2$. Therefore we take 
some $\hat{x}\in\cU^{(1)}_{1}$ and look at the regularity of $\gamma _1$ near $(\hat{x}; \hat{x})$. 
According to Proposition~\ref{prop:mod-C-infini} and Remark~\ref{r:choix-densité}, we have \eqref{eq:mod-C-infini} and, since $N=2$, $\cD$ is a singleton and $c$ is the identity map on $\cD$. Thus, on some neighbourhood $\cV$ of $\hat{x}$, $\gamma _1$ plus a smooth map is given by the function $\gamma : \cV^2\dans\C$ defined by 
\begin{equation}\label{eq:terme-régu-lim}
 \gamma (x; x')\ =\ \int_{\R^3}\, |x\, -\, y|\; \overline{\varphi}(x ; y)\; |x'\, -\, y|\; \varphi(x '; y)\; \chi (y)\, dy\comma 
\end{equation}
where $\varphi :\cV^2\dans\C$ is the sum of a power series and $\chi\in\rC_c^\infty(\R^3; \R^+)$ such that, $\chi =1$ near $\hat{x}$ and its support is contained in $\cV$. 
\begin{remark}\label{r:régu-rho}
 If we set $x=x'$ in \eqref{eq:terme-régu-lim}, we get a squared norm $|x-y|^2$ which is smooth w.r.t. $x$. Standard derivations under the integral sign show that $x\donne \gamma _1(x; x)$ is smooth. When $x\neq x'$, the norm terms are only continuous. It is natural to expect non-smoothness in this case. 
\end{remark}

We start with a lower bound on the regularity of $\gamma$.
\begin{proposition}\label{prop:régu-minimale}
Let $\hat{x}\in\cU^{(1)}_{1}=\R^3\setminus\{R_1;\, \cdots\, ;\, R_L\}$.
Let $n$ be the relevant valuation of the the second decomposition \eqref{eq:décomp-électrons} of Theorem~\ref{th:décomp-anal} near $(\hat{x}; \hat{x})$ in the sense of Definition~\ref{def:valuation}. Near such a point $(\hat{x}; \hat{x})$, the function $\gamma$ {\rm (}see \eqref{eq:terme-régu-lim}{\rm )} belongs to the class $\rC^{n+1}$. 
\end{proposition}
\Pf On the neighbourhood $\cV^2$ of $(\hat{x}; \hat{x})$, we can write $|x-y|\, \overline{\varphi}(x ; y)=|x-y|\overline{\tilde\varphi}(x-y; x+y)$, for some real analytic function $\tilde\varphi$ on some neighbourhood $\cU$ of $(0; 2\hat{x})$ (cf. Definition~\ref{def:valuation}). By Lemma~\ref{lm:régu-valuation}, for fixed $y$, the function $x\donne|x-y|\tilde\varphi (x-y; x+y)$ belongs to the class $\rC^n$ near $\hat{x}$. By standard derivation under the integral sign in \eqref{eq:terme-régu-lim}, we see that the function $\gamma$ belongs to the class $\rC^n$ near $(\hat{x}; \hat{x})$. By Lemma~\ref{lm:régu-valuation}, any partial derivative of order $n+1$ of the function $x\donne|x-y|\tilde\varphi (x-y; x+y)$ is bounded, thus $y$-integrable on the compact support of $\chi$. By Lebesgue's derivation theorem under the integral sign, the function $\gamma$ does belong to the class $\rC^{n+1}$ near the point $(\hat{x}; \hat{x})$.
$\cqfd$

\begin{remark}\label{r:régu-minimale}
 A careful inspection of the proof of Proposition~\ref{prop:régu-minimale} allows us to claim that the functions $x\donne\gamma (x; \hat{x})$ and $x\donne\gamma (\hat{x}; x)$ belong to the class $\rC^{2n+2}$ near $\hat{x}$. Roughly speaking, this can be seen as follows: we differentiate $(n+1)$ times w.r.t. $x$ under the integral sign in \eqref{eq:terme-régu-lim} with $x'=\hat{x}$; then we make the change of variables $y'=y-x$; this allows us to differentiate again $(n+1)$ times w.r.t. $x$ under the integral sign.
\end{remark}
To reveal the limited regularity of $\gamma$, we use the Fourier transform of $\gamma$ times some cut-off function that localizes near $(\hat{x}; \hat{x})$ (in the spirit of the wave front set, cf. \cite{h1} and \eqref{eq:décroiss-poly-illimitée} below). This is based on the 
elementary
\begin{lemma}\label{lm:inté-parties}
Let $d\in\N^\ast$ and $k\in\N$. Let $g: \R^d\dans\C$ be a compactly supported, continuous function and denote by $F_g$ its Fourier transform. Given a real $r$, we denote by $E(r)$ the integer part of $r$, that is the biggest integer less or equal to $r$. 
\begin{enumerate}
 \item If the function $g$ belongs to the class $\rC^k$, then there exists $C>0$ such that, for all $\xi\in\R^d$ with $|\xi|\geq 1$, 
 \[\bigl|F_g(\xi)\bigr|\ \leq\ C\; |\xi|^{-k}\period\]
 \item Assume that the function $F_g$ belongs to the class $\rC^0$ and satisfies, for some real $r>E(r)\geq 0$, for all $\xi\in\R^d$ with $|\xi|\geq 1$, 
 \[\bigl|F_g(\xi)\bigr|\ \leq\ C\; |\xi|^{-r-d}\period\]
 Then $g$ belongs to the class $\rC^{E(r)}$. 
\end{enumerate}
\end{lemma}
\Pf Let $\xi\in\R^d$ with $|\xi|\geq 1$. We have 
\begin{equation}\label{eq:fourier-g}
 F_g(\xi)\ =\ \int_{\R^d}\, e^{-i\, \xi\cdot x}\; g(x)\, dx\period
\end{equation}
If $g$ belongs to the class $\rC^k$, we can integrate by parts $k$ times in \eqref{eq:fourier-g}, thanks to the identity 
\[i\, \frac{\xi}{|\xi|^2}\, \cdot\, \nabla_x\, e^{-i\, \xi\cdot x}\ =\ e^{-i\, \xi\cdot x}\period\]
This leads to 
\[F_g(\xi)\ =\ |\xi|^{-k}\, \int_{\R^d}\, e^{-i\, \xi\cdot x}\; g_k(x)\, dx\]
for some compactly supported, continuous function $g_k$. Since the latter integral is bounded w.r.t. $\xi$, we obtain point 1. \\
Under the assumptions of point 2, the function $F_g$ is integrable on $\R^d$. Thus, by Fourier inversion formula, we have, for $x\in\R^d$, 
\begin{equation}\label{eq:fourier-inv-g}
 g(x)\ =\ \int_{\R^d}\, e^{i\, \xi\cdot x}\; F_g(\xi)\, d\xi\period
\end{equation}
By assumption, the partial derivatives of $(x; \xi)\donne e^{i\, \xi\cdot x}\; F_g(\xi)$ w.r.t. $x$ up to order $E(r)$ are $\xi$-integrable thus, by Lebesgue's derivation theorem, 
we can continuously differentiate $E(r)$ times under the integral sign in \eqref{eq:fourier-inv-g} yielding the $\rC^{E(r)}$ regularity for $g$. $\cqfd$

Let $\chi _0\in\rC_c^\infty(\R^3; \R)$ such that $\chi _0=1$ near $\hat{x}$ and $\chi _0\chi =\chi _0$. For fixed $y$ in the support $S_\chi$ of $\chi$, the function 
\[\R^6\, \ni\, (x; x')\ \donne\ \chi _0(x)\; |x\, -\, y|\; \overline{\varphi}(x ; y)\; \chi _0(x')\; |x'\, -\, y|\; \varphi(x '; y)\]
is continuous and compactly supported. Then, its (usual) Fourier transform is the map $\R^6\ni (\xi ; \xi ')\donne F_y(\xi ; \xi ')$ defined by 
\[F_y(\xi ; \xi ')\ =\ \int_{\R^6}\, e^{-i(\xi\cdot x\, +\, \xi '\cdot x')}\, \chi _0(x)\; |x\, -\, y|\; \overline{\varphi}(x ; y)\; \chi _0(x')\; |x'\, -\, y|\; \varphi(x '; y)\, dx\, dx '\comma\]
where $\xi\cdot x$ denotes the usual, scalar product of $\xi\in\R^3$ and $x\in\R^3$. 
By the Fubini theorem, the map $\R^6\ni (\xi ; \xi ')\donne F(\xi ; \xi ')$, defined by 
\begin{align}\label{eq:transf-fourier}
 &\ F(\xi ; \xi ')\\
 \ =&\ \int_{\R^9}\, e^{-i(\xi\cdot x\, +\, \xi '\cdot x')}\, \chi _0(x)\; |x\, -\, y|\; \overline{\varphi}(x ; y)\; \chi _0(x')\; |x'\, -\, y|\; \varphi(x '; y)\; \chi (y)\; dx\, dx '\, dy\comma\nonumber
\end{align}
is the Fourier transform of the map $\gamma_0:\, \R^6\dans\C$ defined by $\gamma _0(x; x')=\gamma (x; x')\chi _0(x)\chi _0(x')$, which is a localized version of $\gamma$. For fixed $y$, we make the change of variables $\tilde{x}=x-y$ and $\tilde{x}'=x'-y$ in \eqref{eq:transf-fourier}, and we rename for simplicity $(\tilde{x}; \tilde{x}')$ by $(x; x')$. We have 
\begin{align}
 &\ F(\xi ; \xi ')\label{eq:transf-fourier-2}\\
 =&\ \int_{\R^9}\, e^{-i(\xi\cdot x+\xi'\cdot x')}\; e^{-i(\xi+\xi')\cdot y}\; \chi _0(x+y)\; |x|\; \overline{\varphi}(x+y;\, y)\; \chi _0(x'+y)\; |x'|\; \varphi(x'+y;\, y)\nonumber\\
 &\hspace{1cm} \chi (y)\;  dx\, dx'\, dy\period\nonumber
\end{align}
Let $\epsilon>0$. We consider $(\xi ; \xi')$ such that $|\xi+\xi'|\geq\epsilon |(\xi ; \xi')|$ and $|(\xi ; \xi')|\geq 1$. Using that 
\[i\, \frac{\xi+\xi'}{|\xi+\xi'|^2}\, \cdot\, \nabla_y\, e^{-i(\xi+\xi')\cdot y}\ =\ e^{-i(\xi+\xi')\cdot y}\]
and the fact that the integrand in \eqref{eq:transf-fourier-2} is a smooth function of $y$, we get by integration by parts w.r.t. $y$ that, for any $q\in\N$, there exists $C_{q; \epsilon}>0$ such that, for all $(\xi ; \xi')$ with $|\xi+\xi'|\geq\epsilon |(\xi ; \xi')|$ and $|(\xi ; \xi')|\geq 1$, 
\begin{equation}\label{eq:décroiss-rapide}
\bigl|F(\xi ; \xi ')\bigr|\ \leq \ C_{q; \epsilon}\, \bigl|(\xi ; \xi')\bigr|^{-q}\period
\end{equation}
We expect that such a large $|(\xi ; \xi')|$ behaviour does not hold true when $\xi+\xi'=0$. To study this point, we take $\omega$ in the unit sphere $\SSS^2$ of $\R^3$, that is $\omega\in\R^3$ with $|\omega|=1$, and $\lambda\geq 1$. We rewrite \eqref{eq:transf-fourier-2}
for $(\xi ; \xi')=(-\lambda\omega; \lambda\omega)$:
\begin{align*}
 &\ F(-\lambda \omega ;\, \lambda\omega)\\
 =&\ \int_{\R^9}\, e^{\lambda i\, \omega\cdot (x-x')}\, \chi _0(x+y)\; |x|\; \overline{\varphi}(x+y;\, y)\; \chi _0(x'+y)\; |x'|\; \varphi(x'+y;\, y)\; \chi (y)\;  dx\, dx'\, dy\period
\end{align*}
Using the function $\tilde\varphi$ introduced in Definition~\ref{def:valuation}, we rewrite this as 
\begin{align}
 F(-\lambda \omega ;\, \lambda\omega)\
 =&\ \int_{\R^9}\, e^{\lambda i\, \omega\cdot (x-x')}\, |x|\; \overline{\tilde\varphi}(x;\, x+2y)\; |x'|\; \tilde\varphi(x';\, x'+2y)\nonumber\\
 &\hspace{1.5cm}\chi _0(x+y)\; \chi _0(x'+y)
 \; \chi (y)\; \tilde{\chi}\bigl(|x|\bigr)\; \tilde{\chi}\bigl(|x'|\bigr)\;  dx\, dx'\, dy\comma\label{eq:F-omega}
\end{align}
where $\tilde\chi\in\rC_c^\infty(\R)$ is such that, for all $x\in\R^3$ and $y\in S_\chi$, $\tilde\chi (|x|)=1$ if $\chi _0(x+y)\neq 0$. We observe that $\tilde\chi =1$ near $0$. From Proposition~\ref{prop:terme-non-nul}, we know that the relevant valuation of the second decomposition \eqref{eq:décomp-électrons} of Theorem~\ref{th:décomp-anal} near $(\hat{x}; \hat{x})$ is an integer $n\in\N$. This means that this integer $n$ is the valuation of $\tilde\varphi$ w.r.t. its first $3$-dimensional variable. We claim that 
\begin{equation}\label{eq:valuation}
n\ \mbox{is also the valuation of the map}\ x\donne \tilde\varphi (x;\, x+2y)\comma\; \mbox{for almost all}\ y\in S_\chi\comma
\end{equation}
and we refer to the Appendix for a proof.\\
Let us choose an integer $m>2(n+4)$ (this requirement will be explained later). 
In particular, we can write, for $(x ; y)$ near $(0; \hat{x})$, 
\begin{equation}\label{eq:dév-tilde-varphi}
\tilde\varphi (x;\, x+2y)\ =\ \sum _{n\leq |\alpha |<m}\, \tilde{a}_\alpha (y)\; x^\alpha\, +\, \sum _{|\alpha |=m}\, \hat{\varphi}_{\alpha}(x;\, y)\; x^\alpha\comma 
\end{equation}
where the functions $\hat{\varphi}_{\alpha}$ are real analytic near $(0; \hat{x})$ and the functions $\tilde{a}_\alpha$ are real analytic near $\hat{x}$. By definition of $n$, the functions $\tilde{a}_\alpha$ for $|\alpha |=n$ are not all zero. \\
We also write a Taylor formula for $\chi _0$ at fixed $y$ with exact remainder as an integral: 
\[\chi _0(x+y)\ =\ \sum _{|\delta |<m-n}\, \frac{\chi_0^{(\delta)}(y)}{\delta !}\; x^\delta \, +\, \sum _{|\delta |=m-n}\, \hat{\chi}_{\delta}(x;\, y)\; x^\delta\comma\]
where the functions $\hat{\chi}_{\delta}$ are smooth near $(0; \hat{x})$. Using the above formulae, we expand the product $\overline{\tilde\varphi}(x;\, x+2y)\, \chi _0(x+y)$ as 
\[\overline{\tilde\varphi}(\tilde{x};\, \tilde{x}+2y)\, \chi _0(\tilde{x}+y)\ =\ \sum _{n\leq |\alpha |<m}\, \overline{a}_\alpha (y)\; \tilde{x}^\alpha\, +\, \sum _{m\leq |\delta |\leq 2m-n}\, \overline{r}_\delta(\tilde{x}; y)\; \tilde{x}^\delta\comma\]
for some smooth functions $a_\alpha$ and $r_\delta$. 
We observe that, for $|\alpha |=n$, $a_\alpha (y)=\tilde{a}_\alpha(y)\chi _0(y)$. Of course, we may replace in the above expansion the variable $x$ by the variable $x'$. Now we insert those expansions into the formula \eqref{eq:F-omega}. We observe that, 
by the Fubini theorem, the resulting terms in \eqref{eq:F-omega} contain $3$-dimensional integrals of the form
\begin{equation}\label{eq:int-typique}
 \int_{\R^3}\, e^{-\lambda i\, \tilde\omega\cdot x}\; |x|\; \;  x^\delta\; \tilde{\chi}(|x|)\; \bigl(a(y)\, +\, r(x ;\, y)\bigr)\; dx\comma
\end{equation}
for some smooth functions $r$ and $a$, $\delta\in\N^3$, and $\tilde\omega =\pm\omega$. Using the fact that $x\donne |x|\, x^\delta\, \tilde{\chi}(|x|)$ belongs to the class $\rC^{|\delta|}$, by Lemma~\ref{lm:régu-valuation}, and using 
\begin{equation}\label{eq:exp}
 e^{-i\lambda\, \tilde\omega\cdot x}\ =\ i\lambda ^{-1}(\tilde\omega\cdot\nabla _x)e^{-i\lambda\, \tilde\omega\cdot x}\comma
\end{equation}
we see by integrations by parts that integrals of the type \eqref{eq:int-typique} are $O\bigl(\lambda^{-|\delta|}\bigr)$, uniformly w.r.t. $y\in S_\chi$. This allows us to rearrange \eqref{eq:F-omega} as 
\begin{align}
 &F(-\lambda \omega ;\, \lambda\omega)\, +\, O\bigl(\lambda^{-m}\bigr)\label{eq:dev-reste-lambda-1}\\
 \ =&\ \sum _{n\leq |\alpha |<m\atop n\leq |\alpha '|<m}\, \int_{\R^9}\, e^{\lambda i\, \omega\cdot (x-x')}\, |x|\; x^\alpha\; \tilde{\chi}\bigl(|x|\bigr)\; |x'|\; (x')^{\alpha '}\; \tilde{\chi}\bigl(|x'|\bigr)\; \overline{a}_{\alpha}(y)\; a_{\alpha '}(y)\; \chi (y)\;  dx\, dx'\, dy\period\nonumber
\end{align}
For $\alpha\in\N^3$, we denote by $F_\alpha$ the Fourier transform of the continuous, compactly supported, real-valued function $\R^3\ni x\donne |x|\, x^\alpha\, \tilde{\chi}(|x|)$. Then \eqref{eq:dev-reste-lambda-1} can be written as 
\begin{align}
 &F(-\lambda \omega ;\, \lambda\omega)\nonumber\\
 \ =&\ \sum _{n\leq |\alpha |<m\atop n\leq |\alpha '|<m}\, F_\alpha (-\lambda\omega)\; F_{\alpha '}(\lambda\omega)\; \int_{\R^3}\, 
 \overline{a}_{\alpha}(y)\; a_{\alpha '}(y)\; \chi (y)\, dy\ +\ O\bigl(\lambda^{-m}\bigr)\nonumber\\
 \ =&\ \sum _{n\leq |\alpha |<m\atop n\leq |\alpha '|<m}\, \overline{F_\alpha (\lambda\omega)}\; F_{\alpha '}(\lambda\omega)\; \int_{\R^3}\, 
 \overline{a}_{\alpha}(y)\; a_{\alpha '}(y)\; \chi (y)\, dy\ +\ O\bigl(\lambda^{-m}\bigr)\period\label{eq:dev-reste-lambda-2}
\end{align}
Let us denote by $F_0$ the Fourier transform of the continuous, compactly supported function $\tilde{f} :\, \R^3\dans\C$, defined by $\tilde{f}(x)=|x|\, \tilde{\chi}(|x|)$. The function $F_0$ is smooth. For all $\alpha\in\N^3$, we observe that $F_\alpha$ is actually $(i\partial _\xi)^\alpha F_0$ ($\xi$ being the Fourier variable associated to $x$).\\
To extract from \eqref{eq:dev-reste-lambda-2} the large $\lambda$ asymptotics, we use 
the elementary 

\begin{lemma}\label{lm:fourier-norme-localisée}
Let $f : \R^3\ni x\donne |x|\cdot\tau \bigl(|x|\bigr)$ where $\tau\in\rC_c^\infty(\R)$ such that $\tau =1$ near $0$. Then, its Fourier transform $F_f$ is a smooth function on $\R^3$, which is given, for $\xi\neq 0$, by 
\begin{equation}\label{eq:fourier-norme-localisée}
F_f(\xi)\ =\ \frac{4\pi}{|\xi|}\, \int_0^{+\infty}\, \tau (r)\; r^2\; \sin (r|\xi|)\, dr\period
\end{equation}
It has the following behaviour at infinity: 
\begin{equation}\label{eq:fourier-norme-localisée-symbole-1}
 \forall\, \alpha\in\N^3\comma\ \exists\, C_\alpha>0\, ;\ \forall\, \xi\in \R^3\setminus\{0\}\comma\ \bigl|\partial^\alpha F_f(\xi)\bigr|\ \leq\ C_\alpha\;  |\xi |^{-4-|\alpha|}\period
\end{equation}
Furthermore, there exists a smooth function $G :\R^3\setminus\{0\}\dans\R$ such that, for $\xi\neq 0$, $F_f(\xi )=-8\pi |\xi |^{-4}+G(\xi )$ and such that 
\begin{equation}\label{eq:fourier-norme-localisée-symbole-2}
 \forall\, k\in\ncg 5; +\infty\ncg\comma\, \forall\, \alpha\in\N^3\comma\ \exists\, C_{k; \alpha}>0\, ;\ \forall\, \xi\in \R^3\setminus\{0\}\comma\ \bigl|\partial^\alpha G(\xi)\bigr|\ \leq\ C_{k; \alpha}\;  |\xi |^{-k-|\alpha|}\period
\end{equation}
\end{lemma}
\Pf See the Appendix. $\cqfd$

By Lemma~\ref{lm:fourier-norme-localisée} with $\tau=\tilde{\chi}$ and the fact that $m\geq 2n+9$, we derive from \eqref{eq:dev-reste-lambda-2} the estimates $F(-\lambda \omega ;\, \lambda\omega)=O(\lambda^{-2n-8})$ and 
\begin{align}
 F(-\lambda \omega ;\, \lambda\omega)\ =&\ \sum _{|\alpha |=n\atop |\alpha '|=n}\, \overline{F_\alpha (\lambda\omega)}\; F_{\alpha '}(\lambda\omega)\;  \int_{\R^3}\, \overline{\tilde{a}}_{\alpha}(y)\; \tilde{a}_{\alpha '}(y)\; \chi _0^2(y)\, dy
 \ +\ O\bigl(\lambda^{-2n-9}\bigr)\comma\label{eq:ordre-lambda-0}\\
 \ =&\ \int_{\R^3}\, \; \bigl|f_n(y; \lambda)\bigr|^2\; \chi _0^2(y)\, dy\ +\ O\bigl(\lambda^{-2n-9}\bigr)\comma\label{eq:ordre-lambda-1}
\end{align}
where 
\[f_n(y; \lambda)\ :=\ \sum _{|\alpha |=n}\, F_\alpha (\lambda\omega)\; \tilde{a}_{\alpha}(y)\ =\ \sum _{|\alpha |=n}\, \tilde{a}_{\alpha}(y)\, \bigl((i\partial_\xi)^\alpha F_0\bigr)_{|\xi =\lambda\omega}\period\]
Using the decomposition of $F_f$, \eqref{eq:fourier-norme-localisée-symbole-1}, and \eqref{eq:fourier-norme-localisée-symbole-2} in Lemma~\ref{lm:fourier-norme-localisée} with $\tau=\tilde{\chi}$ and the homogeneity of the partial derivatives of the function $|\cdot|^{-4}$, we get the expansion 
\begin{align}
 &\ F(-\lambda \omega ;\, \lambda\omega)\nonumber\\
 \ =&\ (8\pi )^2\, \int_{\R^3}\, \biggl|\sum _{|\alpha |=n}\, \tilde{a}_{\alpha}(y)\, \bigl((i\partial _\xi)^\alpha\, |\cdot|^{-4}\bigr)_{|\xi =\lambda\omega}\biggr|^2\, \chi _0^2(y)\, dy\ +\ O\bigl(\lambda^{-2n-9}\bigr)\nonumber\\
 \ =&\ \lambda^{-2n-8}\, (8\pi )^2\, \int_{\R^3}\, \biggl|\sum _{|\alpha |=n}\, \tilde{a}_{\alpha}(y)\, \bigl((i\partial _\xi)^\alpha\, |\cdot|^{-4}\bigr)_{|\xi =\omega}\biggr|^2\, \chi _0^2(y)\, dy\ +\ O\bigl(\lambda^{-2n-9}\bigr)\period\label{eq:ordre-lambda-2}
\end{align}
Now, \eqref{eq:ordre-lambda-2} is really a large $\lambda$ asymptotics of $F(-\lambda \omega ;\, \lambda\omega)$ if the first term on its r.h.s. has size $\lambda^{-2n-8}$ in the sense that there exist positive constants $c, c'$ such that $\lambda^{2n+8}$ times this first term belongs to $[c; c']$. This is precisely the case when the integral in \eqref{eq:ordre-lambda-2} is nonzero. 
We observe that this holds true if and only if the first term on the r.h.s. of \eqref{eq:ordre-lambda-1} 
has size $\lambda^{-2n-8}$ too. \\
It turns out that \eqref{eq:ordre-lambda-1} and \eqref{eq:ordre-lambda-2} are both a large $\lambda$ asymptotics of $F(-\lambda \omega ;\, \lambda\omega)$ for some $\omega\in\SSS^2$, as shown in 
\begin{lemma}\label{lm:vraie-asymptotique}
 Let $\hat{x}\in\cU^{(1)}_{1}=\R^3\setminus\{R_1;\, \cdots\, ;\, R_L\}$. For any $\omega\in\SSS^2$, the first term on the r.h.s of \eqref{eq:ordre-lambda-1} has size $\lambda^{-2n-8}$ if and only if the first term on the r.h.s of \eqref{eq:ordre-lambda-2} has size $\lambda^{-2n-8}$. Furthermore, there exists $\omega\in\SSS^2$ such that the first term on the r.h.s of \eqref{eq:ordre-lambda-2} has size $\lambda^{-2n-8}$, that is, the integral there is nonzero. 
\end{lemma}
\Pf We already proved the first statement. \\
Assume that the integral in \eqref{eq:ordre-lambda-2} is zero for any $\omega\in\SSS^2$. Since it is homogeneous as a function of $\omega$, it is actually zero for any $\omega\in\R^3\setminus\{0\}$. By the properties of $\chi_0$ and the continuity of the map 
\[y\ \donne\ \sum _{|\alpha |=n}\, \tilde{a}_{\alpha}(y)\, \bigl((i\partial _\xi)^\alpha\, |\cdot|^{-4}\bigr)_{|\xi =\omega}\comma\]
the latter function is identically zero on the support of $\chi_0$, for any fixed $\omega\in\R^3\setminus\{0\}$. Since the functions $\tilde{a}_{\alpha}$ are continuous on the support of $\chi_0$, they all are bounded there. By Lemma~\ref{lm:fourier-norme-localisée}, we can find some $C>0$ such that, for all $\omega\in\SSS^2$, for all $\lambda\geq 1$, and for all $y$ in the support of $\chi_0$, 
\begin{equation}\label{eq:borne-lambda}
\bigl|f_n(y; \lambda)\bigr|\ =\ \left|\sum _{|\alpha |=n}\, \tilde{a}_{\alpha}\bigl(y\bigr)\, \bigl((i\partial _\xi)^\alpha\, F_0\bigr)_{|\xi =\lambda\omega}\right|\ \leq\ C\; \lambda^{-5-n}\period
\end{equation}
Let us fix $y$ in the support of $\chi_0$. Recall that $F_0$ is the Fourier transform of the function $\tilde{f}:\, x\, \donne\, |x|\, \tilde{\chi}(|x|)$. The term $f_n(y; \lambda)$, the norm of which is estimated in \eqref{eq:borne-lambda}, is the Fourier transform of the function $h_y:\, \R^3\dans\C$ given by $h_y(x)\, =\, |x|\, \varphi _y(x)\, \tilde{\chi}(|x|)$, 
where $\varphi _y$ is the real analytic function defined by 
\[\R^3\ni x\ \donne\ \varphi _y(x)\ :=\ \sum _{|\alpha |=n}\, \tilde{a}_{\alpha}\bigl(y\bigr)\, x^\alpha\period\]
By \eqref{eq:borne-lambda} and Lemma~\ref{lm:inté-parties} (with $d=3$ and $r\in ]1+n; 2+n[$), $h_y$ belongs to the class $\rC^{n+1}$. This is also true on a small ball $\tilde\cB$ that is centered at $0$, is independent of $y$, and on which $\tilde\chi =1$. By Lemma~\ref{lm:régu-valuation}, the valuation of $\varphi _y$ has to be larger than $n$, unless $\varphi _y$ is identically zero. By the definition of $\varphi _y$, however, the valuation of $\varphi_y$ is less or equal to $n$, thus $\varphi _y$ is identically zero on $\tilde\cB$. \\
Therefore, for $(x; y)\in\tilde\cB\times S_{\chi _0}$, $\varphi_y(x)=0$ and, in particular, for $|\alpha|=n$ and $y\in S_{\chi _0}$, $\tilde{a}_\alpha (y)=0$. This contradicts the statement made just after \eqref{eq:dév-tilde-varphi} on the functions $\tilde{a}_\alpha$.\\
Therefore, we have proven that, for some $\omega\in\SSS^2$, the first term on the r.h.s of \eqref{eq:ordre-lambda-2} has size $\lambda^{-2n-8}$. $\cqfd$

\begin{proposition}\label{prop:régu-limitée}
Let $\hat{x}\in\cU^{(1)}_{1}=\R^3\setminus\{R_1;\, \cdots\, ;\, R_L\}$. Let $n$ be the relevant valuation of the the second decomposition \eqref{eq:décomp-électrons} of Theorem~\ref{th:décomp-anal} near $(\hat{x}; \hat{x})$ in the sense of Definition~\ref{def:valuation}. In a vicinity of such a point $(\hat{x}; \hat{x})$, the function $\gamma$ {\rm (}see \eqref{eq:terme-régu-lim}{\rm )} does not belong to the class $\rC^{2n+9}$. 
\end{proposition}
\Pf Assume that the function $\gamma$ belongs to the class $\rC^{2n+9}$ near $(\hat{x}; \hat{x})$. In particular, for some neighbourhood $\cV_0$ of $\hat{x}$ such that $\chi =1$ on $\cV_0$, the restriction of $\gamma$ to $\cV_0^2$ belongs to the class $\rC^{2n+9}$. Let $\chi _0\in\rC_c^\infty(\R^3; \R)$ such that $\chi _0=1$ near $\hat{x}$ and the support $S_{\chi _0}$ of $\chi _0$ is included in $\cV_0$. For such a cut-off $\chi _0$, $\gamma_0$ belongs to the class $\rC^{2n+9}$. Thus, by Lemma~\ref{lm:inté-parties}, we have a bound 
\begin{equation}\label{eq:borne-fourier}
 \exists\, C>0\, ;\ \forall\, (\xi\, ;\, \xi')\in\R^6\setminus\{(0;0)\}\comma\hspace{.4cm}\bigl|F(\xi ; \xi ')\bigr|\ \leq\ C\; \bigl|(\xi\, ;\, \xi')\bigr|^{-2n-9}\period
\end{equation}
But, at the same time, for such a cut-off $\chi _0$, we know from the above computation that there exists some $\omega\in\SSS^2$ such that, for $\lambda\geq 1$, $F(-\lambda \omega ;\, \lambda\omega)$ is exactly of order $\lambda^{-2n-8}$ (in the sense that \eqref{eq:ordre-lambda-1} and \eqref{eq:ordre-lambda-2} are true expansions for large $\lambda$). This contradicts \eqref{eq:borne-fourier}. Therefore, $\gamma$ cannot belong to the class $\rC^{2n+9}$ near $(\hat{x}; \hat{x})$. $\cqfd$

To describe a little bit more how $\gamma_1$ is not smooth, we try to compute its wave front set ``above a vicinity'' of $(\hat{x}; \hat{x})$ ($\hat{x}$ being as in Proposition~\ref{prop:régu-limitée}). We recall first the definition of the wave front set (see \cite{h2}, p. 254). \\
The wave front set ${\rm WF}(\gamma _1)$ of $\gamma _1$ is a (possibly empty) subset of $\R^6\times (\R^6\setminus\{(0; 0)\})$, that is conical in the second variable. A point $(x; x'; \xi; \xi ')\in (\R^6\times \R^6\setminus\{(0; 0)\})$ does not belong to the wave front set of $\gamma _1$ if there exists some $\tau\in\rC_c^\infty(\R^6; \C)$ such that $\tau (x; x')\neq 0$ and a conical neighbourhood $\Gamma$ of $(\xi; \xi ')$ such that the Fourier transform $F_{\tau\gamma_1}$ of $\tau\gamma _1$ satisfies 
\begin{equation}\label{eq:décroiss-poly-illimitée}
 \forall\, p\in\N\comma\ \exists\, C_p>0\, ;\ \forall\, (\eta ; \eta ')\in \Gamma\setminus\{(0; 0)\}\comma\ \bigl|F_{\tau\gamma_1}(\eta ; \eta ')\bigr|\ \leq\ C_p\; \Bigl(1\, +\, \bigl|(\eta ; \eta ')\bigr|\Bigr)^{-p}\comma
\end{equation}
where $|\cdot|$ denotes the euclidian norm on $\R^6$. Let $P$ be the projection 
\[\bigl(\R^6\times \R^6\setminus\{(0; 0)\}\bigr)\ni \bigl(x; x'; \xi; \xi '\bigr)\ \donne\ \bigl(x; x'\bigr)\period\]
It is well-known that $P({\rm WF}(\gamma _1))$ is precisely the singular support of $\gamma_1$, that is the complement of the largest open set in $\R^6$ on which $\gamma _1$ is smooth (see \cite{h2}, p. 254). By Theorem~\ref{th:régu-anal}, we know that the singular support of $\gamma _1$ must be a subset of the complement of $\cU^{(2)}_1$. This complement contains the diagonal $D$ of $\R^6$. Proposition~\ref{prop:régu-limitée} tells us that each $(\hat{x}; \hat{x})$, with $\hat{x}\not\in\{R_1;\, \cdots ;\, R_L\}$, belongs to the singular support of $\gamma_1$. Since the singular support is always closed, the singular support of $\gamma _1$ contains $D$. \\
If $s$ is a smooth function on $\R^6$ then, for $\tau\in\rC_c^\infty(\R^6; \C)$, \eqref{eq:décroiss-poly-illimitée} holds true with $F_{\tau\gamma_1}$ replaced by $F_{\tau s}$ and $\Gamma$ replaced by $\R^6$. 
Since $\gamma _1=\gamma +s$ near $(\hat{x}; \hat{x})$ for such a smooth function $s$ (cf. Proposition~\ref{prop:mod-C-infini}), $\gamma _1$ and $\gamma$ have the same wave front set ``above'' $(\hat{x}; \hat{x})$. \\
Given $\omega\in\SSS^2$, it seems intuitively that \eqref{eq:décroiss-poly-illimitée} for $\gamma$ and for $(\xi ; \xi ')=(-\omega ; \omega)$ should not hold true when the integral in \eqref{eq:ordre-lambda-2} is nonzero. This is not obvious since $\tau$ may be different from the cut-off function $\tau _0 : \R^6\ni (x; x')\donne\chi _0(x)\chi _0(x')$ used in \eqref{eq:transf-fourier}, but it is true as shown in
\begin{lemma}\label{lm:manque-décroissance}
 Let $\hat{x}\in\cU^{(1)}_{1}=\R^3\setminus\{R_1;\, \cdots\, ;\, R_L\}$. Take $\omega\in\SSS^2$ such that the integral in \eqref{eq:ordre-lambda-2} is nonzero. Then $(\hat{x}; \hat{x}; -\omega; \omega)\in{\rm WF}(\gamma _1)$. 
\end{lemma}
\Pf See the Appendix. $\cqfd$

We are able to give the following information on the wave front set of $\gamma_1$.
\begin{proposition}\label{prop:front-onde}
Let $\hat{x}_0\in\cU^{(1)}_{1}=\R^3\setminus\{R_1;\, \cdots\, ;\, R_L\}$. Let $\cV_0$ be a neighbourhood of $\hat{x}_0$ such that $\cV_0\subset\cU^{(1)}_{1}$.
\begin{enumerate}
 \item The wave front set ${\rm WF}(\gamma _1)$ of $\gamma _1$ above $\cV_0^2$ is included in the ``conormal set of the diagonal'' above $\cV_0^2$, that is 
\begin{equation}\label{eq:conormal-diag}
 {\rm WF}(\gamma _1)\, \cap\, \Bigl(\cV_0^2\times\bigl(\R^6\setminus\{(0; 0)\}\bigr)\Bigr)\ \subset\ \bigl\{(\hat{x}; \hat{x}; -\xi; \xi);\; \hat{x}\in\cV_0\comma\; \xi\in\bigl(\R^3\setminus\{0\}\bigr)\bigr\}\period
\end{equation}
 \item Let $\cA$ be the subset of $\SSS^2$ formed by the $\omega\in\SSS^2$ for which the integral in \eqref{eq:ordre-lambda-2} is nonzero.
 We denote by $\overline{\cA}$ the smallest closed subset of $\SSS^2$ that contains $\cA$. Then 
\begin{equation}
\emptyset\ \neq\ \bigl\{(\hat{x}; \hat{x}; -\lambda\omega; \lambda\omega);\; \hat{x}\in\cV_0\comma\; \lambda\in\ ]0; +\infty[\comma\; \omega\in\overline{\cA}\bigr\}\ \subset\ {\rm WF}(\gamma _1)\period\label{eq:mino-front-onde}
 \end{equation}
\end{enumerate}
\end{proposition}
\Pf First, we point out that a wave front set is always closed (cf. \cite{h2}, p. 254). Recall that, on the considered region, ${\rm WF}(\gamma _1)$ and ${\rm WF}(\gamma)$ coïncide. 
\begin{enumerate}
 \item Let $(\hat{x}; \hat{x}; \xi _0; \xi_0')$ with $\hat{x}\in\cU^{(1)}_{1}$ and $\xi_0+\xi_0'\neq 0$. In a small enough, closed,  conical neighbourhood $\Gamma$ of $(\xi _0; \xi_0')$, on which $\xi+\xi'\neq 0$, we get the bound \eqref{eq:décroiss-rapide}. This yields \eqref{eq:décroiss-poly-illimitée} for $\gamma$ and shows that $(\hat{x}; \hat{x}; \xi _0; \xi_0')\not\in WF(\gamma )$. This proves \eqref{eq:conormal-diag}
 \item By Lemma~\ref{lm:vraie-asymptotique} and Lemma~\ref{lm:manque-décroissance}, 
\[\emptyset\ \neq\ \bigl\{(\hat{x}; \hat{x}; -\omega; \omega)\, ;\ \hat{x}\in\cV_0\comma\; \omega\in\cA\bigr\}\ \subset\ {\rm WF}(\gamma)\period\]
Since a wave front set is always closed, 
\[\bigl\{(\hat{x}; \hat{x}; -\omega; \omega)\, ;\ \hat{x}\in\cV_0\comma\; \omega\in\overline{\cA}\bigr\}\ \subset\ {\rm WF}(\gamma )\period\]
Since ${\rm WF}(\gamma )$ is conical in the last two variables, we obtain \eqref{eq:mino-front-onde}. $\cqfd$
\end{enumerate}
The property \eqref{eq:conormal-diag} is true if $\gamma_1$ is the kernel of a pseudodifferential operator, the symbol of which belongs to a certain class of smooth symbols, see Theorem 18.1.16 in \cite{h3} p. 80. Furthermore, (a localized version of) $\gamma_1$ can always be seen as the kernel of a pseudodifferential operator (cf. \cite{h3} p. 69). A natural question arises: does the symbol of the pseudodifferential operator associated to (a localized version of) $\gamma_1$ belong to a class of smooth symbols? By Proposition~\ref{prop:mod-C-infini}, it suffices to consider $\gamma$ instead of $\gamma _1$. 
We gives below a positive answer to this question for the localized version $\gamma_0$ of $\gamma$.

\begin{proposition}\label{prop:symbole}
Let $\hat{x}\in\cU^{(1)}_{1}=\R^3\setminus\{R_1;\, \cdots\, ;\, R_L\}$. Let $n$ be the relevant valuation of the the second decomposition \eqref{eq:décomp-électrons} of Theorem~\ref{th:décomp-anal} near $(\hat{x}; \hat{x})$ in the sense of Definition~\ref{def:valuation}. Let $\chi _0\in\rC_c^\infty(\R^3; \R)$ such that $\chi _0=1$ near $\hat{x}$ and $\chi _0\chi =\chi _0$. Let $a :\, \R^6\dans\C$ be defined by, for  $(x;\, \xi)\in\R^6$, 
\[a(x;\, \xi)\ =\ \int_{\R^3}\, e^{-i\, \xi\cdot t}\; \gamma_0\bigl(x-t/2;\, x+t/2\bigr)\, dt\period\]
Then, the function $a$ is smooth. Moreover
\begin{equation}\label{eq:carac-symbole}
\forall\, (\alpha ; \beta)\in \bigl(\N^3\bigr)^2\comma\ \sup_{(x;\, \xi)\in \R^3\times\R^3}\, \bigl(1\, +\, |\xi|\bigr)^{n+|\beta|}\, \Bigl|\bigl(\partial _x^\alpha\partial_\xi^\beta a\bigr)(x;\, \xi)\Bigr|\ <\ +\infty\period
\end{equation}
\end{proposition}
\Pf We first show that the function $a$ is smooth. Let $(x;\, \xi)\in\R^6$. Using \eqref{eq:terme-régu-lim}, the change of variables $y'=x-y$, and the function $\tilde\varphi$ from Definition~\ref{def:valuation}, we have 
\begin{align}
 a(x;\, \xi)\ =\ &\int_{\R^6}\, e^{-i\, \xi\cdot t}\; \bigl|x\, -\, t/2\, -\, y\bigr|\; \overline{\varphi}\bigl(x -t/2; y\bigr)\; \bigl|x\, +\, t/2\, -\, y\bigr|\; \varphi\bigl(x +t/2; y\bigr)\nonumber\\
 &\hspace{1.4cm} \chi _0\bigl(x-t/2\bigr)\, \chi _0\bigl(x+t/2\bigr)\, \chi (y)\, dy\, dt\nonumber\\
 \ =\ &\int_{\R^6}\, e^{-i\, \xi\cdot t}\; \bigl|y'\, -\, t/2\bigr|\; \bigl|y'\, +\, t/2\bigr|\; \overline{\varphi}\bigl(x -t/2; x-y'\bigr)\; \; \varphi\bigl(x +t/2; x-y'\bigr)\nonumber\\
 &\hspace{1.4cm} \chi _0\bigl(x-t/2\bigr)\, \chi _0\bigl(x+t/2\bigr)\, \chi (x-y')\, dy'\, dt\nonumber\\
\ =\  &\int_{\R^6}\, e^{-i\, \xi\cdot t}\; \bigl|y\, -\, t/2\bigr|\; \overline{\tilde\varphi}\bigl(y-t/2; 2x-t/2-y\bigr)\; 
\nonumber\\
&\hspace{1.4cm} \bigl|y\, +\, t/2\bigr|\;  \tilde\varphi\bigl(y+t/2; 2x+t/2-y\bigr)\nonumber\\
 &\hspace{1.4cm} \chi _0\bigl(x-t/2\bigr)\, \chi _0\bigl(x+t/2\bigr)\, \chi (x-y)\, dy\, dt\period\label{eq:formule-a}
\end{align}
Now, we can use standard derivations under the integral sign in both $x$ and $\xi$, yielding the smoothness of $a$, since the cut-off functions $\chi$ and $\chi_0$ are smooth and compactly supported. \\
We observe also that the integration in \eqref{eq:formule-a} takes place in the $x$-dependent, compactly supported region 
\[\cR_x\ :=\ \bigl\{(t; y)\in\R^6\, ;\ (x-t/2)\in S_{\chi_0}\comma\, (x+t/2)\in S_{\chi_0}\comma\, (x-y)\in S_\chi\bigr\}\comma\]
that is sent to a $x$-independent, compact subset $K$ of $S_{\chi_0}\times S_\chi$ by the $x$-dependent maps $(t; y)\donne (y-t/2; 2x-t/2/-y)$ and $(t; y)\donne (y+t/2; 2x+t/2/-y)$. By Lemma~\ref{lm:régu-valuation}, we know that the function $g : S_{\chi_0}\times S_\chi\ni (x; y)\donne |x|\tilde\varphi (x; y)$ belongs to the class $\rC^n$. This implies that the integrand in \eqref{eq:formule-a} belongs also, for fixed $y$, to the class $\rC^n$ in the variable $t$. Using, for nonzero $\xi$, 
\[i\, \frac{\xi}{|\xi|^2}\, \cdot\, \nabla_t\, e^{-i\, \xi\cdot t}\ =\ e^{-i\, \xi\cdot t}\]
and integrations by parts in \eqref{eq:formule-a}, we get, for $|\xi|\geq 1$, 
\[a(x;\, \xi)\ =\ |\xi|^{-n}\, \int_{\cR_x}\, e^{-i\, \xi\cdot t}\; g_n\bigl(t; y; x; \xi/|\xi|\bigr)\, dt\, dy\comma\]
for some continuous function $g_n$ satisfying the following property: there exists $C>0$, depending on a finite number of partial derivatives of the functions $\chi _0$, $\chi$, and $g$ and on the compact $K$, such that, for $x\in\R^3$ and $\xi\in\R^6\setminus\{0\}$, 
\[\int_{\cR_x}\, \bigl|g_n\bigl(t; y; x; \xi/|\xi|\bigr)\bigr|\, dt\, dy\, \leq\, C\period\]
This yields the inequality in \eqref{eq:carac-symbole} when $\alpha =\beta =0$. In a similar way, we obtain also this inequality when $\beta=0$. Observe that $(\partial _\xi^\beta a)(x; \xi)$ is just given by \eqref{eq:formule-a} with the integrand replaced by its $\partial _\xi^\beta$-partial derivative. Precisely, 
\begin{align}
 (\partial _\xi^\beta a)(x;\, \xi)\ =\ &(-i)^{|\beta|}\int_{\R^6}\, e^{-i\, \xi\cdot t}\; t^\beta\; \bigl|y\, -\, t/2\bigr|\; \overline{\tilde\varphi}\bigl(y-t/2; 2x-t/2-y\bigr)\; 
\nonumber\\
&\hspace{2.2cm}\bigl|y\, +\, t/2\bigr|\;  \tilde\varphi\bigl(y+t/2; 2x+t/2-y\bigr)\nonumber\\
 &\hspace{2.2cm}\chi _0\bigl(x-t/2\bigr)\, \chi _0\bigl(x+t/2\bigr)\, \chi (x-y)\, dy\, dt\period\label{eq:formule-dériv-a}
\end{align}
Writing $t=(t/2-y)+(t/2+y)$, we see that the valuation of the integrand as a function of $y-t/2$ (resp. $y+t/2$) is now $n+|\beta|$, at least. 
By the above argument, we obtain the inequality in \eqref{eq:carac-symbole} when $\alpha =0$. Repeating those arguments with $a$ replaced by $(\partial _x^\alpha a)$, we complete the proof of \eqref{eq:carac-symbole}. $\cqfd$

\begin{remark}\label{r:gamma1}
We provide here several remarks on the previous results. 
\begin{enumerate}
 \item Propositions~\ref{prop:mod-C-infini} and~\ref{prop:régu-limitée} give Theorem~\ref{th:non-lisse} in the two-electron case. 
 \item The proof of Proposition~\ref{prop:régu-limitée} gives a rather crude estimation of the regularity of $\gamma$. This is due to the fact that, in general, bounds on the Fourier transform of a function {\rm (}distribution{\rm)} are not a precise tool to determine the exact regularity of that function. 
 \item In Proposition~\ref{prop:symbole}, the function $a$ is the symbol of the Weyl pseudodifferential operator associated to $\gamma_0$ {\rm (}see the Weyl calculus in \cite{h3} p. 150{\rm)}. The property \eqref{eq:carac-symbole} means that the function $a$ belongs to the class $S^{-n}$ on $\R^6$, in the sense of Definition 18.1.1 p. 65 in \cite{h3}. 
 \item We refer to \cite{h3}, p. 80, for the notion of ``conormal set of the diagonal''. Proposition~\ref{prop:symbole} together with Theorem 18.1.16 on p. 80 in \cite{h3} imply \eqref{eq:conormal-diag}. 
 \item In Propostion~\ref{prop:front-onde}, we believe that $\overline{\cA}=\SSS^2$. In that case, the wave front set of $\gamma _1$ above $\cV_0$ would be precisely the conormal of the diagonal above $\cV_0$, that is, \eqref{eq:conormal-diag} would be an equality. 
 \item It is quite remarkable that a localized version of $\gamma_1$ defines a pseudodifferential operator, the {\rm (}Weyl{\rm)}  symbol of which belongs to a standard class of smooth symbols. This means that one can more or less include $\gamma_1$ into some quite regular pseudodifferential calculus. More precisely, the density matrix $\gamma_1$ is the kernel of a pseudodifferential operator on the open set $\cU^{(1)}_{1}=\R^3\setminus\{R_1;\, \cdots\, ;\, R_L\}$, the symbol of which belongs to the class $S_{\rm loc}^{-n}(\cU^{(1)}_{1}\times\R^3)$ {\rm (}cf. \cite{h3}, p. 83{\rm)}. 
\end{enumerate}
\end{remark}
%

\section{The general case.}
\label{s:general}
\setcounter{equation}{0}

In this section, we perform the same tasks as in Section~\ref{s:2-e} in the general case. We shall see that we can follow the arguments of Section~\ref{s:2-e} with minor changes. We consider the Hamiltonian \eqref{eq:hamiltonien} with arbitrary $N\geq 2$ and $L\geq 1$.

We consider a point $(\hat{\ux}; \hat{\ux}')\in (\cU^{(1)}_{N-1}\times\cU^{(1)}_{N-1})\, \cap\, \cC _{N-1}^{(2)}$, where the sets $\cU^{(1)}_{N-1}$ and $\cC _{N-1}^{(2)}$ are given by \eqref{eq:sans-coll-k} and \eqref{eq:collisions-extérieures}, respectively. Like in Section~\ref{s:2-e}, the starting point of our analysis is formula \eqref{eq:mod-C-infini} in Proposition~\ref{prop:mod-C-infini}. 
For $j\in\cD$ and $(\ux ; \ux ')\in\cV\times\cV'$, we set 
\begin{equation}\label{eq:partie-irrég-j}
 \gamma_{N-1}^{(j)}(\ux ;\, \ux')\ :=\ \int_{\R^3}\, |x_j\, -\, y|\, \overline{\varphi_j}(\ux ; y)\, |x_{c(j)}'\, -\, y|\, \varphi_{c(j)}'(\ux '; y)\chi _j(y)\, dy\period
\end{equation}
We observe that $\gamma_{N-1}^{(j)}$ has the same structure as $\gamma$ in \eqref{eq:terme-régu-lim}: the variable $(x_j; x_{c(j)}')$ plays the rôle of the variable $(x; x')$ in $\gamma$. The first one varies in a vicinity of $(\hat{x}_j; \hat{x}_j)$ (since $\hat{x}_{c(j)}'=\hat{x}_j$) and the second one stays in a neighbourhood of $(\hat{x}; \hat{x})$. We possibly have additional $x$ and $x'$ variables since $\ux =(x_j;\, (x_k)_{k\neq j})$ and $\ux '=(x_{c(j)}';\, (x_k')_{k\neq c(j)})$. The function $\varphi _j$ plays the rôle of the real analytic function $\varphi _2$ of the decomposition \eqref{eq:décomp-électrons} with $k=N$ near $\hat{\uz}:=(\hat{\ux}; \hat{x}_j)$, while the function $\varphi _{c(j)}$ plays the rôle of the real analytic function $\varphi _2$ of the decomposition \eqref{eq:décomp-électrons} with $k=N$ near $\hat{\uz}:=(\hat{\ux}'; \hat{x}_{c(j)}')$ (cf. the proof of Proposition~\ref{prop:mod-C-infini}). According to Remark~\ref{r:choix-densité}, we may assume that the $\varphi _j$ and $\varphi _{c(j)}$ are the sum of a power series. Let $n_j$ be the relevant valuation associated to the decomposition \eqref{eq:décomp-électrons} near $\hat{\uz}:=(\hat{\ux}; \hat{x}_j)$ and let $n_j'$ be the relevant valuation associated to the decomposition \eqref{eq:décomp-électrons} near $\hat{\uz}:=(\hat{\ux}'; \hat{x}_{c(j)}')=(\hat{\ux}'; \hat{x}_j)$. We immediately see that we can follow the arguments of the proof of Proposition~\ref{prop:régu-minimale} to get 
\begin{proposition}\label{prop:régu-minimale-géné}
Let $(\hat{\ux}; \hat{\ux}')\in (\cU^{(1)}_{N-1}\times\cU^{(1)}_{N-1})\, \cap\, \cC _{N-1}^{(2)}$. Let $j\in\cD$. 
Then the function $\gamma_{N-1}^{(j)}$ {\rm (}see \eqref{eq:partie-irrég-j}{\rm )} belongs to the class $\rC^{\bar{n}_j+1}$ near $(\hat{\ux}; \hat{\ux}')$, where $\bar{n}_j=\min (n_j; n_j')$. In particular, the function $\gamma_{N-1}$ belongs to the class $\rC^{\bar{n}+1}$ near $(\hat{\ux}; \hat{\ux}')$, where $\bar{n}=\min\{\bar{n}_j; j\in\cD\}$.
\end{proposition}
\Pf Similar to the proof of Proposition~\ref{prop:régu-minimale}. $\cqfd$

Now, we try to show that the density $\gamma_{N-1}$ has a limited regularity near such a point $(\hat{\ux}; \hat{\ux}')\in (\cU^{(1)}_{N-1}\times\cU^{(1)}_{N-1})\, \cap\, \cC _{N-1}^{(2)}$. This question actually reduces to the same question for the terms $\gamma_{N-1}^{(j)}$, defined in \eqref{eq:partie-irrég-j} for $j\in\cD$. Indeed, for such a $j$, the term $\gamma_{N-1}^{(j)}$ has an unlimited regularity w.r.t. to $x_k$ for $k\neq j$ and to $x_k'$ for $k\neq c(j)$. Thus the other terms $\gamma_{N-1}^{(j')}$, for $j'\in\cD\setminus\{j\}$, cannot cancel the nonsmoothness of $\gamma_{N-1}^{(j)}$ w.r.t. the variables $(x_{j}; x_{c(j)}')$, if such nonsmoothness is true. \\
Let $j\in\cD$. We introduce a localized version of $\gamma_{N-1}^{(j)}$ and consider its Fourier transform. Denoting by $F$ this Fourier transform, we essentially have the formula \eqref{eq:transf-fourier}: to be precise, the first function $\chi _0$ is replaced by an appropriate cut-off function in the variable $\ux$, the second one is replaced by an appropriate cut-off function in the variable $\ux'$, the integration takes place on $(\R^{3(N-1)})^2$, the Fourier variables 
\[\uxi\ =\ (\xi _1;\, \cdots ;\, \xi_{N-1})\hspace{.4cm}\mbox{and}\hspace{.4cm}\uxi'\ =\ (\xi _1';\, \cdots ;\, \xi_{N-1}')\] 
belong to $\R^{3(N-1)}$, and, more importantly, the first function $\varphi$ and the second one are replaced by the functions $\varphi _j$ and $\varphi _{c(j)}'$, respectively, that appear in \eqref{eq:partie-irrég-j}. A careful inspection of the arguments developed in Section~\ref{s:2-e} shows that we can follow them with $\gamma$ replaced by $\gamma_{N-1}^{(j)}$ and get, instead of \eqref{eq:ordre-lambda-0}, a similar expansion for $F(\uxi ; \uxi')$, where $\lambda\geq 1$, $\omega\in\SSS^2$, $\xi_k=0$ if $k\neq j$, $\xi_k'=0$ if $k\neq c(j)$, and $\xi _j=\lambda\omega=-\xi _{c(j)}$. We do not see why this formula should be a true large $\lambda$ asymptotics, except when we require that $\hat{x}=\hat{x}'$. Indeed, in this case, the real analytic functions $\varphi _j$ and $\varphi _j'$ are equal (cf. Remark~\ref{r:irrégu-gamma}) and we can use some nonnegativity as in the proof of Lemma~\ref{lm:vraie-asymptotique} to get the result. \\
We expect that this is also true for many $(\hat{\ux}; \hat{\ux}')$, but that, for some others, the first term of the expansion is zero but not the following one. Because of the real analyticity involved in the problem, we believe that, in general, $F(\uxi ; \uxi')$ is of some finite order in $\lambda$ above or equal to $n_j+n_j'$.

\begin{remark}\label{r:cas-géné-diag}
Let $(\hat{\ux}; \hat{\ux}')\in (\cU^{(1)}_{N-1}\times\cU^{(1)}_{N-1})\, \cap\, \cC _{N-1}^{(2)}$. The above exploration encourages us to believe that each $\gamma_{N-1}^{(j)}$ does have a limited regularity near $(\hat{\ux}; \hat{\ux}')$, and so does the density $\gamma_{N-1}$. Only in the ``diagonal'' case $\hat{\ux}=\hat{\ux}'$, we see how to get an upper bound on those regularities. 
\end{remark}

Now, we prove Theorem~\ref{th:non-lisse} in its full generality. 

\Pfof{of Theorem~\ref{th:non-lisse}}Let us take $\hat{\ux}\in\cU^{(1)}_{N-1}$. By Proposition~\ref{prop:mod-C-infini} and Remark~\ref{r:choix-densité}, there is a neighbourhood $\cV$ of $\hat{\ux}$ such that, on $\cV^2:=\cV\times\cV$, $\gamma_{N-1}=s+\sum _{j=1}^{N-1}\gamma_{N-1}^{(j)}$ and, for $(\ux ;\, \ux')\in\cV^2$ and $j\in\ncg 1; N-1\ncd$, 
\begin{equation}\label{eq:partie-irrég-j-bis}
 \gamma_{N-1}^{(j)}(\ux ;\, \ux')\ :=\ \int_{\R^3}\, |x_j\, -\, y|\, \overline{\varphi_j}(\ux ; y)\, |x_j'\, -\, y|\, \varphi_j(\ux '; y)\chi _j(y)\, dy\period
\end{equation}
Let $j\in\ncg 1; N-1\ncd$ and denote by $\tilde\varphi _j$ the real analytic function defined in the second case of Definition~\ref{def:valuation} with $k=N$, when $\varphi _2$ is replaced by $\varphi_j$. Recall that we may assume that 
each $\tilde\varphi _j$ is the sum of a power series and that $n_j$ is the valuation of $\tilde\varphi _j$ w.r.t. its $j$th variable. \\
We first show that $\gamma_{N-1}^{(j)}$ does not belong to the class $\rC^{2n_j+9}$ near $(\hat{\ux}; \hat{\ux})$. \\
By definition of $n_j$, we can write on $\cV\times S_{\chi _j}$, 
\begin{align*}
 \varphi _j(\ux ;\, y)=\ &\tilde\varphi _j\bigl(x_1;\, \cdots; x_{j-1};\, x_j-y;\, x_{j+1}; \cdots;\, x_{N-1};\, x_j+y\bigr)\\
 =\ &\sum_{\alpha _j\in\N^3}\, \varphi _{\alpha _j}\bigl(x_1;\, \cdots; x_{j-1};\, x_{j+1}; \cdots;\, x_{N-1}; \, x_j+y\bigr)\; (x_j\, -\, y)^{\alpha _j}\comma
\end{align*}
for some real analytic functions $\varphi _{\alpha _j}$ with $\alpha _j\in\N^3$. Since the functions $\varphi _{\alpha _j}$ with $\alpha _j\in\N^3$ and $|\alpha _j|=n_j$ are not all zero, there exists 
\[\bigl(x_1^0;\, \cdots; x_{j-1}^0;\, x_{j+1}^0; \cdots;\, x_{N-1}^0\bigr)\, \in\, \R^{3(N-2)}\]
and a neighbourhood $\cV_j$ of $\hat{x}_j$ such that, for $x_j\in\cV_j$ and $y\in S_{\chi _j}$, 
\[\bigl(x_1;\, \cdots; x_{j-1};\, x_j;\, x_{j+1}; \cdots;\, x_{N-1};\, y\bigr)\, \in\, \cV\]
and such that one of the functions 
\[y'\donne \varphi _{\alpha _j}\bigl(x_1^0;\, \cdots; x_{j-1}^0;\, x_{j+1}^0; \cdots;\, x_{N-1}^0; \, y'\bigr)\comma\]
for $\alpha _j\in\N^3$ with $|\alpha _j|=n_j$, is nonzero. In particular, for $y\in S_{\chi _j}$ and $x$ near zero in $\R^3$, 
\begin{align*}
&\tilde\varphi _j\bigl(x_1^0;\, \cdots; x_{j-1}^0;\, x;\, x_{j+1}^0; \cdots;\, x_{N-1}^0;\, x+2y\bigr)\\
=\ &\sum _{|\alpha |=n_j}\, \tilde{a}^{(j)}_\alpha (y)\; x^\alpha\, +\, \sum _{|\alpha |>n_j}\, \hat{\varphi}_{\alpha}^{(j)}(\tilde{x};\, y)\; x^\alpha\comma
\end{align*}
where one of the real analytic functions $\tilde{a}^{(j)}_\alpha$, for $\alpha\in\N^3$ with $|\alpha |=n_j$, is nonzero, and the functions $\hat{\varphi}_{\alpha}^{(j)}$ are real analytic near $(0; \hat{x}_j)$. \\
Now, we observe that the map 
\[(x; \, x')\donne\gamma_{N-1}^{(j)}\bigl(x_1^0;\, \cdots; x_{j-1}^0;\, x;\, x_{j+1}^0; \cdots;\, x_{N-1}^0;\, x_1^0;\, \cdots; x_{j-1}^0;\, x';\, x_{j+1}^0; \cdots;\, x_{N-1}^0\bigr)\comma\]
defined near $(\hat{x}_j; \hat{x}_j)$, has exactly the same structure as $\gamma$ in \eqref{eq:terme-régu-lim}. The proof of Proposition~\ref{prop:régu-limitée} shows that this function cannot belong the class $\rC^{2n_j+9}$ near $(\hat{x}_j; \hat{x}_j)$. Therefore, the full function $\gamma_{N-1}^{(j)}$ cannot belong the class $\rC^{2n_j+9}$ near $(\hat{\ux}; \hat{\ux})$. \\
Let $k\in\ncg 1; N-1\ncd\setminus\{j\}$. By standard derivation w.r.t. $x_j$ under the integral sign, we see that the map 
\[(x; \, x')\donne\gamma_{N-1}^{(k)}\bigl(x_1^0;\, \cdots; x_{j-1}^0;\, x;\, x_{j+1}^0; \cdots;\, x_{N-1}^0;\, x_1^0;\, \cdots; x_{j-1}^0;\, x';\, x_{j+1}^0; \cdots;\, x_{N-1}^0\bigr)\]
is actually smooth near $(\hat{x}_j; \hat{x}_j)$. Using \eqref{eq:mod-C-infini}, we conclude that the 
\[(x; \, x')\donne\gamma_{N-1}\bigl(x_1^0;\, \cdots; x_{j-1}^0;\, x;\, x_{j+1}^0; \cdots;\, x_{N-1}^0;\, x_1^0;\, \cdots; x_{j-1}^0;\, x';\, x_{j+1}^0; \cdots;\, x_{N-1}^0\bigr)\]
cannot belong the class $\rC^{2n_j+9}$ near $(\hat{x}_j; \hat{x}_j)$. In particular, the full density $\gamma_{N-1}$ cannot belong the class $\rC^{2n_j+9}$ near $(\hat{\ux}; \hat{\ux})$. 
$\cqfd$

We end this section with some words on the extension to the general case of the other results in Section~\ref{s:2-e}, namely  Propositions~\ref{prop:front-onde} and~\ref{prop:symbole}. We use the decomposition \eqref{eq:mod-C-infini} of $\gamma_{N-1}$ obtained in  Proposition~\ref{prop:mod-C-infini}. Let $j\in\cD$. 
Using integration by parts, we see as in the proof of point 1 in Proposition~\ref{prop:front-onde} that the wave front set of $\gamma_{N-1}^{(j)}$ above a point $(\hat{\ux}; \hat{\ux}')\in (\cU^{(1)}_{N-1}\times\cU^{(1)}_{N-1})\, \cap\, \cC _{N-1}^{(2)}$ is included in 
\[\bigl\{(\hat{\ux}; \hat{\ux}'; \uxi; \uxi ')\in \R^{6(N-1)}\, ;\ \xi_j+\xi_j'\, =\, 0\bigr\}\period\]
We deduce from the decomposition \eqref{eq:mod-C-infini} of the density $\gamma_{N-1}$ that its wave front set above such a point $(\hat{\ux}; \hat{\ux}')$ is included in 
\[\bigl\{(\hat{\ux}; \hat{\ux}'; \uxi; \uxi ')\in \R^{6(N-1)}\, ;\ \forall\, j\in\cD\comma\ \xi_j+\xi_j'\, =\, 0\bigr\}\period\]
For such a point $(\hat{\ux}; \hat{\ux}')\in (\cU^{(1)}_{N-1}\times\cU^{(1)}_{N-1})\, \cap\, \cC _{N-1}^{(2)}$, we also get an extension of Proposition~\ref{prop:symbole} on each term $\gamma_{N-1}^{(j)}$, for $j\in\cD$. \\
Let $j\in\cD$. We take a cut-off function $\chi _0\in\rC_c^\infty(\R^{3(N-1)})$ (resp. $\chi _0'\in\rC_c^\infty(\R^{3(N-1)})$) that localizes near $\hat{\ux}$ (resp. $\hat{\ux}'$). Similar to the function $a$ in Proposition~\ref{prop:symbole}, we introduce 
\[a^{(j)}(\ux ; \uxi)\ :=\ \int_{\R^{3(N-1)}}\, e^{-i\, \xi\cdot t}\; \chi _0(\ux-t/2)\, \gamma_{N-1}^{(j)}\bigl(\ux-t/2;\, \ux+t/2\bigr)\, \chi _0'(\ux '+t/2)\, dt\period\]
Using the arguments of the proof of Proposition~\ref{prop:symbole}, we see that this function $a^{(j)}$ is smooth and that 
\[\forall\, (\alpha ; \beta)\in \bigl(\N^{3(N-1)}\bigr)^2\comma\ \sup_{(\ux;\, \uxi)\in \R^{3(N-1)}\times\R^{3(N-1)}}\, \bigl(1\, +\, |\uxi|\bigr)^{n_j+|\beta|}\, \Bigl|\bigl(\partial _\ux^\alpha\partial_\uxi^\beta a^{(j)}\bigr)(\ux;\, \uxi)\Bigr|\ <\ +\infty\period\]
This means that $a^{(j)}$ belongs to the symbol class $S^{-n_j}$ on $\R^{3(N-1)}\times\R^{3(N-1)}$ (cf. Definition 18.1.1 p. 65 in \cite{h3}). A careful inspection even shows that one has the property:
\begin{align*}
 &\forall\, m\in\N\comma\ \forall\, (\alpha ; \beta)\in \bigl(\N^{3(N-1)}\bigr)^2\comma\\
 &\sup_{(\ux;\, \uxi)\in \R^{3(N-1)}\times\R^{3(N-1)}}\, \bigl(1\, +\, |\uxi^{(j)}|\bigr)^{m+|\beta^{(j)}|}\, \bigl(1\, +\, |\xi_j|\bigr)^{n_j+|\beta_j|}\, \Bigl|\bigl(\partial _\ux^\alpha\partial_\uxi^\beta a^{(j)}\bigr)(\ux;\, \uxi)\Bigr|\ <\ +\infty\comma
\end{align*}
where we have used $\uxi =(\uxi ^{(j)}; \xi _j)$ and $\beta =(\beta^{(j)}; \beta_j)$. Roughly speaking, we have an unlimited ``decay'' w.r.t. the variable $\uxi ^{(j)}$ and a limited one w.r.t. the variable $\xi _j$.

\vspace{.5cm}
\begin{appendices}{\bf \Large Appendix.}

\renewcommand{\theequation}{{\rm A}.\arabic{equation}} 

\appendixtitleon
\appendixtitletocon


\vspace{.5cm}

For completeness, we provide in this Appendix a proof for the more or less known results stated in the Lemmata~\ref{lm:régu-valuation}, \ref{lm:fourier-norme-localisée}, and \ref{lm:manque-décroissance}. We also prove the important claim \eqref{eq:valuation}. 

First of all, we need to recall a basic property of the differential calculus. Let $O$ be an open set of $\R^p$, $p\in\N^\ast$, let $f, g : O\dans\C$ be two functions in the class $\rC^k$, for some $k\in\N$. Then, for any $\alpha\in\N^d$ with $|\alpha |\leq k$, we have, on $O$,
\begin{equation}\label{eq:binomial-formula}
\partial^\alpha\bigl(fg\bigr) (x)\ =\ \sum _{\gamma\in\N^p\atop \gamma\leq\alpha}\, \mathfrak{b}_\gamma^\alpha\, \bigl(\partial^{\gamma}f\bigr)(x)\, \bigl(\partial^{\alpha -\gamma}g\bigr)(x)\ =\ \sum _{\gamma\in\N^p\atop \gamma\leq\alpha}\, \mathfrak{b}_\gamma^\alpha\, \bigl(\partial^{\alpha -\gamma}f\bigr)(x)\, \bigl(\partial^{\gamma}g\bigr)(x)\comma
\end{equation}
where the binomial coefficients $\mathfrak{b}_\gamma^\alpha$ are given by 
\begin{equation}\label{eq:binomial}
 \mathfrak{b}_\gamma^\alpha\ :=\ \frac{\alpha !}{\bigl((\alpha -\gamma)!\bigr)\cdot \bigl(\gamma !\bigr)}\period
\end{equation}
We also use the following variant. Let $\omega\in\R^p$. Then, for an integer $m\leq k$ and for $x\in O$, 
\begin{equation}\label{eq:binomial-formula-bis}
 (\omega\cdot\nabla _x)^m(fg)(x)\ =\ \sum _{p=0}^m\, \mathfrak{c}_p^m\, (\omega\cdot\nabla _x)^pf(x)\, (\omega\cdot\nabla _x)^{m-p}g(x)\comma
\end{equation}
where the binomial coefficients $\mathfrak{c}_p^k$ are given by 
\begin{equation}\label{eq:binomial-bis}
 \mathfrak{c}_p^m\ :=\ \frac{m!}{\bigl((m-p)!\bigr)\cdot \bigl(p!\bigr)}\period
\end{equation}
The binomial coefficients \eqref{eq:binomial} also enter into the following multivariable binomial formula: For $\alpha\in\N^d$, $(x; y)\in (\R^d)^2$, 
\[(x\, +\, y)^\alpha\ =\ \sum _{\gamma\in\N^p\atop \gamma\leq\alpha}\, \mathfrak{b}_\gamma^\alpha\, x^\gamma\, y^{\alpha - \gamma}\period\]

\Pfof{the claim \eqref{eq:valuation}}By the choice of $\chi$ and of the neighbourhood $\cV$, we know that $\varphi (x; y)$ is the sum of a power series on $\cV\times S_\chi$. Setting, on $\cV\times S_\chi$, $u=x-y\in\R^3$ and $v=x+y\in\R^3$, the variable $u$ lives in some neighbourhood $\cW_0$ of $0$ and the variable $v$ lives in some neighbourhood $\cW$ of $2\hat{x}$. According to Definition~\ref{def:valuation}, we can write, for $(u; v)\in\cW_0\times\cW$, 
\begin{equation}\label{eq:dév-tilde-phi}
 \tilde\varphi (u; v)\ =\ \sum _{\alpha\in\N^3\atop |\alpha|\geq n}\, \varphi _\alpha (v)\, u^\alpha\comma
\end{equation}
a convergent series, where the functions $\varphi _\alpha$ are also convergent power series of the form 
\[\varphi _\alpha (v)\ =\ \sum _{\beta\in\N^3}\, c(\alpha; \beta)\, (v\, -\, 2\hat{x})^\beta\period\]
Let $y\in S_\chi$ and $x\in\cW_0$. We have $x+2y\in\cW$. Let $\alpha\in\N^3$ with $|\alpha|\geq n$. 
Using the multivariable newtonian binomial formula, we get 
\begin{align*}
 \varphi _\alpha (x+2y)\ =&\ \sum _{\beta '\in\N^3}\, c(\alpha; \beta ')\, (x\, +\, 2y\, -\, 2\hat{x})^{\beta '}\ =\ \sum _{\beta '\in\N^3}\, c(\alpha; \beta ')\, \sum _{\beta\leq\beta '}\, \mathfrak{b}_\beta^{\beta '}\, x^\beta\, (2y\, -\, 2\hat{x})^{\beta '-\beta}\\
 \ =&\ \sum _{\beta\in\N^3}\, x^\beta\, \sum _{\beta '\geq\beta}\, c(\alpha; \beta ')\, \mathfrak{b}_\beta ^{\beta '}\,(2y\, -\, 2\hat{x})^{\beta '-\beta}\ =\ \sum _{\beta\in\N^3}\, x^\beta\, \varphi _{\alpha ; \beta}(y)\comma
\end{align*}
where 
\[\varphi _{\alpha ; \beta}(y)\ =\ \sum _{\beta '\geq\beta}\, c(\alpha; \beta ')\, \mathfrak{b}_\beta ^{\beta '}\,(2y\, -\, 2\hat{x})^{\beta '-\beta}\period\]
Therefore, 
\[\tilde\varphi (x ; x+2y)\ =\ \sum_{\alpha\in\N^3\atop |\alpha|\geq n}\, x^\alpha\, \sum _{\beta\in\N^3}\, x^\beta\, \varphi _{\alpha ; \beta}(y)\ =\ \sum_{\alpha\in\N^3\atop |\alpha|=n}\, x^\alpha\, \varphi _{\alpha ; 0}(y)\, +\, \sum_{\alpha\in\N^3\atop |\alpha|>n}\, x^\alpha\, \tilde\varphi _\alpha(y)\comma\]
for appropriate power series $\tilde\varphi _\alpha(y)$. In particular, the valuation of the map $x\donne\tilde\varphi (x ; x+2y)$ is greater or equal to $n$. \\
If, for some $\alpha$, the function $\varphi _{\alpha ; 0}$ is identically zero, all the coefficients $c(\alpha; \beta)$ are zero and $\varphi _\alpha$ is also identically zero. By definition of the valuation $n$, the functions $\varphi _\alpha$ with $|\alpha |=n$ cannot all be identically zero. Thus, there exists
such a multiindex $\alpha$ for which $\varphi _{\alpha ; 0}$ is not identically zero. As a real analytic map, $\varphi _{\alpha ; 0}$ takes, for almost all $y$, a nonzero value $\varphi _{\alpha ; 0}(y)$. Therefore, for almost all $y\in S_\chi$, the valuation of the map $x\donne\tilde\varphi (x ; x+2y)$ is exactly $n$, yielding the claim \eqref{eq:valuation}. $\cqfd$

For the proof of Lemma~\ref{lm:régu-valuation}, we need some basic facts. Let $\omega\in\R^3$ with $|\omega|=1$. Notice that 
$(\omega\cdot\nabla _x)(\omega\cdot x)=1$, on $\R^3$. We consider $P_0\in\R[X]$ and $P_1\in\R[X]$ the polynomials with real coefficients given by $P_0(X)=1$ and $P_1(X)=X$. Setting $\hat{x}:=x/|x|$, for $x\in\R^3\setminus\{0\}$, we observe that, on $\R^3\setminus\{0\}$, $(\omega\cdot\nabla _x)(|x|)=\omega\cdot\hat{x}=|x|^0P_1(\omega\cdot\hat{x})$, $(\omega\cdot\nabla _x)^0(|x|)=|x|P_0(\omega\cdot\hat{x})$, and 
\[(\omega\cdot\nabla _x)(|x|^{-1})\ =\ -\, \frac{\omega\cdot\hat{x}}{|x|^2}\period\]
Still on $\R^3\setminus\{0\}$,
\[(\omega\cdot\nabla _x)^2(|x|)\ =\ (\omega\cdot\nabla _x)\, (\omega\cdot\hat{x})\ =\ \frac{1}{|x|}\, -\, \frac{(\omega\cdot x)(\omega\cdot\hat{x})}{|x|^2}\ =\ \frac{1}{|x|}\, P_2\bigl(\omega\cdot\hat{x}\bigr)\comma\]
where $P_2(X)=1-X^2$. By a straightforward induction, we show that, for any integer $k$ with $k\geq 2$, $(\omega\cdot\nabla _x)^k(|x|)=|x|^{1-k}P_k(\omega\cdot\hat{x})$ on $\R^3\setminus\{0\}$, where $P_k\in\R[X]$ satisfies the following properties: the degree of $P_k$ is $k$, the polynomial $P_2$ is a divisor of $P_k$, and the dominant coefficient of $P_k$ is 
\[(-1)^{k+1}\, \prod_{j=1}^{k-1}\, (2j-1)\period\]
Let $p\in\N$. For $\alpha\in\N^3$ with $|\alpha|=p$, let $H_\alpha :\, \SSS^2\ni\omega\donne\omega^{\alpha}\in\R$. We show that the maps $H_\alpha$, for $\alpha\in\N^3$ with $|\alpha|=p$, are linearly independent. \\
Assume that there are complex numbers $c_\alpha$, $\alpha\in\N^3$ with $|\alpha|=p$, such that the map $\displaystyle\sum _{|\alpha|=p}\, c_\alpha\, H_\alpha$ is zero identically. Then, for all $x\in\R^3\setminus\{0\}$, 
\[0\ =\ \sum _{|\alpha|=p}\, c_\alpha\, \left(\frac{x}{|x|}\right)^\alpha\ =\ |x|^{-|\alpha|}\, \sum _{|\alpha|=p}\, c_\alpha\, x^\alpha\]
This shows that the continuous map $\R^3\ni x\donne\displaystyle\sum _{|\alpha|=p}\, c_\alpha\, x^\alpha$ is zero identically. It is well known that the homogeneous polynomials $\R^3\ni x\donne x^\alpha$, for $\alpha\in\N^3$ with $|\alpha|=p$, are linearly independent. Thus, $c_\alpha=0$ for each $\alpha$. This proves that the $H_\alpha$, for $\alpha\in\N^3$ with $|\alpha|=p$, are linearly independent. 

We split the proof of Lemma~\ref{lm:régu-valuation} in two steps. 

\Pfof{Lemma~\ref{lm:régu-valuation}, special case}For $\alpha\in\N^3$, let $N_\alpha : \R^3\dans\R$ be the function defined by $N_\alpha (x)=|x|x^{\alpha}$. Here we show that the function $N_\alpha$ does not belong to the class $\rC^{|\alpha|+1}$, that is a part of the statement of Lemma~\ref{lm:régu-valuation} for a particular case. \\
By induction, we first show that, for $\alpha\in\N^3$ and $k\in\N$, we have, on $\R^3$, 
\[(\omega\cdot\nabla _x)^k(x^\alpha)\ =\ \sum _{\gamma\leq\alpha\atop|\gamma|=k}\, \frac{\alpha !}{(\alpha -\gamma)!}\; \omega^\gamma\; x^{\alpha -\gamma}\period\]
Let $\alpha\in\N^3$ and $m\in\N$. Making use of the previous facts and of \eqref{eq:binomial-formula-bis}, we can compute, on 
$\R^3\setminus\{0\}$, $(\omega\cdot\nabla _x)^m(N_\alpha)$. In particular, we get, for $m=|\alpha|+1$ and $x\in\R^3\setminus\{0\}$, 
\begin{align}\label{eq:dérivée-ordre-spécial}
 \bigl((\omega\cdot\nabla _x)^mN_\alpha\bigr)(x)\ =\ &\alpha !\, \bigl(|\alpha|+1\bigr)\, P_1\bigl(\omega\cdot\hat{x}\bigr)\\
 &\ +\, \sum _{\gamma<\alpha}\frac{\alpha !}{(\alpha -\gamma)!}\; \mathfrak{b}_{|\gamma|}^{|\alpha|+1}\; P_{|\alpha|+1-|\gamma|}\bigl(\omega\cdot\hat{x}\bigr)\; \omega^\gamma\; x^{\alpha -\gamma}\; |x|^{|\gamma|-|\alpha|}\nonumber
\end{align}
(cf. \eqref{eq:binomial} with $p=1$) and observe that it is a function of $\hat{x}=x/|x|$ only. For $\epsilon\in\{-1; 1\}$, we compute its values at $\hat{x}=\epsilon\omega$. 
Since $\omega\cdot\hat{x}=\epsilon$ and $P_2$ is a divisor of $P_k$ (for $k\geq 2)$, we get from \eqref{eq:dérivée-ordre-spécial} that 
\[\bigl((\omega\cdot\nabla _x)^mN_\alpha\bigr)_{|\hat{x}=\epsilon\omega}\ =\ \alpha !\, \bigl(|\alpha|+1\bigr)\, \omega^\alpha\, \epsilon\ +\ 0\period\]
In particular, this yields the existence and the value of the following
\[\lim _{t\to 0\atop t>0}\, \bigl((\omega\cdot\nabla _x)^mN_\alpha\bigr)_{|x=t\epsilon\omega}\ =\ \alpha !\, \bigl(|\alpha|+1\bigr)\, \omega^\alpha\, \epsilon\period\]
If the function $N_\alpha$ would belong to the class $\rC^{|\alpha|+1}$, then we should have the equalities 
\begin{align*}
 \alpha !\, \bigl(|\alpha|+1\bigr)\, \omega^\alpha\ =\ &\lim _{t\to 0\atop t>0}\, \bigl((\omega\cdot\nabla _x)^mN_\alpha\bigr)_{|x=t\omega}\\
 \ =\ &\lim _{t\to 0\atop t>0}\, \bigl((\omega\cdot\nabla _x)^mN_\alpha\bigr)_{|x=-t\omega}\ =\ -\, \alpha !\, \bigl(|\alpha|+1\bigr)\, \omega^\alpha\period
\end{align*}

This would imply $\omega^\alpha=0$, for all $\omega\in\SSS^2$. Contradiction. 
Therefore, the function $N_\alpha$ does not belong to the class $\rC^{|\alpha|+1}$. $\cqfd$

We turn now to the proof of Lemma~\ref{lm:régu-valuation} in its full generality. 

\Pfof{Lemma~\ref{lm:régu-valuation}}Recall that $\cW$ is a bounded neighbourhood of $0$ and $N_\varphi :\cW\dans\C$ is defined by 
$N_\varphi (x)=|x|\varphi (x)$. The function $N_\varphi$ is smooth on $\cW\setminus\{0\}$, as the product of smooth functions on this region. If $\alpha\in\N^3$ with $|\alpha|=q+1$, then, on $\cW\setminus\{0\}$, using the formula \eqref{eq:binomial-formula}, 
\[\bigl(\partial _x^\alpha N_\varphi \bigr)(x)\ =\ \sum _{\gamma\leq\alpha}\, \mathfrak{b}_\gamma^\alpha \bigl(\partial _x^\gamma (|x|)\bigr)(x)\, \bigl(\partial _x^{\alpha -\gamma}\varphi\bigr)(x)\comma
\]
where $\partial _x^{\alpha -\gamma}\varphi$ is identically zero or has valuation $q-|\alpha|+|\gamma|$. Since, for all multiindex $\gamma$ with $\gamma\leq\alpha$, the functions 
$\cW\setminus\{0\}\ni x\donne |x|^{|\gamma|-1}\bigl(\partial _x^\gamma (|x|)\bigr)(x)$ are bounded, so is also $\partial _x^\alpha N_\varphi$ on $\cW\setminus\{0\}$. \\
We prove by induction on $q\in\N$ the property: For any real analytic function $\varphi$, that is defined on a neighbourhood $\cW_0$ of $0$ and the valuation of which is greater or equal to $q$, the function  $N_\varphi$ belongs to the class $\rC^q$ on $\cW_0$ and all the partial derivatives up to the order $q$ of $N_\varphi$ vanish at zero. \\
Since the euclidian norm $|\cdot|$ is continuous and vanishes at zero, the property is valid for $q=0$. Assume that it is true for some $q\in\N$. Let $\varphi$ be a real analytic function $\varphi$ on some neighbourhood $\cW_0$ of $0$ such that its valuation is greater or equal to $q+1$. We can find real analytic functions $\varphi _j :\, \cW_0\dans\C$, for $j\in\ncg 1; 3\ncd$, such that, for $x\in\cW_0$, 
\begin{equation}\label{eq:phi-décomp}
 \varphi(x)\ =\ \sum _{j=1}^3\, x_j\, \cdot\, \varphi _j(x)\period
\end{equation}
For $j\in\ncg 1; 3\ncd$, $\varphi _j$ is either zero identically or its valuation is at least $q$.
By the induction hypothesis, the functions $N_{\varphi _j}: \cW_0\ni x\donne |x|\varphi _j(x)$ belong to the class $\rC^q$ and their partial derivatives up to order $q$ all vanish at zero. Furthermore, by the previous result, all the partial derivative of order $q+1$ of $N_{\varphi _j}$ are bounded near $0$. \\
Using \eqref{eq:phi-décomp}, we see that $N_\varphi$ belongs to the class $\rC^q$. Let $\alpha\in\N^3$ with $|\alpha|\leq q$. We can write, for $x\in\cW_0$, using \eqref{eq:phi-décomp} and \eqref{eq:binomial-formula}, 
\begin{equation}\label{eq:dérivée-partielle-formule}
 \bigl(\partial _x^\alpha N_\varphi\bigr)(x)\ =\ \sum _{j=1}^3\, \sum _{\gamma\leq\alpha}\, \mathfrak{b}_\gamma^\alpha \bigl(\partial _x^\gamma (x_j)\bigr)(x)\, \bigl(\partial _x^{\alpha -\gamma}N_{\varphi _j}\bigr)(x)\period
\end{equation}
We observe that, for $x\in\R^3$, $\bigl(\partial _x^\gamma (x_j)\bigr)(x)=0$ if $|\gamma|>1$ or if $|\gamma|=1$ and $\gamma_j=0$. Thus, the sum in \eqref{eq:dérivée-partielle-formule} simplifies to 
\begin{equation}\label{eq:dérivée-partielle-formule-simple}
 \bigl(\partial _x^\alpha N_\varphi\bigr)(x)\ =\ \sum _{j=1}^3\, \sum _{\gamma\leq\alpha\atop |\gamma|=1=\gamma _j}\, \mathfrak{b}_\gamma^\alpha \bigl(\partial _x^{\alpha -\gamma}N_{\varphi _j}\bigr)(x)\ +\ \sum _{j=1}^3\, x_j\; \bigl(\partial _x^{\alpha}N_{\varphi _j}\bigr)(x)\period
\end{equation}
By the induction hypothesis applied to the $\varphi _j$, we see that $\bigl(\partial _x^\alpha N_\varphi\bigr)(0)=0$. \\
Let $\alpha\in\N^3$ with $|\alpha|=q$. Denote by $(e_1; e_2; e_3)$ the canonical basis of $\R^3$. 
For $t\in\R^\ast$ small enough and $k\in\{1; 2; 3\}$, we consider 
\begin{equation}\label{eq:taux}
 t^{-1}\, \left(\bigl(\partial _x^\alpha N_\varphi\bigr)(te_k)\, -\, \bigl(\partial _x^\alpha N_\varphi\bigr)(0)\right)\ =\ t^{-1}\, \bigl(\partial _x^\alpha N_\varphi\bigr)(te_k)\comma
\end{equation}
insert into \eqref{eq:taux} the formula \eqref{eq:dérivée-partielle-formule-simple}. Observe that the contribution in \eqref{eq:taux} of the last term of \eqref{eq:dérivée-partielle-formule-simple} is
\[t^{-1}\, \sum _{j=1}^3\, t\, \delta_{jk}\; \bigl(\partial _x^{\alpha}N_{\varphi _j}\bigr)(te_k)\ =\   \bigl(\partial _x^{\alpha}N_{\varphi _k}\bigr)(te_k)\period\]
By the induction hypothesis applied to the $\varphi _j$, we see that \eqref{eq:taux} tends to zero, as $t\to 0$. This shows that the partial first derivative w.r.t. $x_k$ at zero of $\partial _x^\alpha N_\varphi$ exists and is zero. Away from zero, this partial derivative is, thanks to \eqref{eq:dérivée-partielle-formule-simple}, given by
\begin{align}
\partial _{x_k}\Bigl(\bigl(\partial _x^\alpha N_\varphi\bigr)\Bigr)(x)\ =\ & \bigl(\partial _x^\alpha N_{\varphi_k}\bigr)(x)\, +\, \sum _{j=1}^3\, x_j\, \bigl(\partial _{x_k}\partial _x^{\alpha}N_{\varphi_j}\bigr)(x)\label{eq:1-de-plus}\\
&\ +\, \sum _{j=1}^3\, \sum _{\gamma\leq\alpha\atop |\gamma|=1=\gamma _j}\, \mathfrak{b}_\gamma^\alpha\, \bigl(\partial _{x_k}\partial _x^{\alpha -\gamma}N_{\varphi_j}\bigr)(x)\period\nonumber
\end{align}
Using again the induction hypothesis and the fact that any partial derivative of $N_{\varphi_j}$ of order $q+1$ is bounded near $0$, for all $j\in\{1; 2; 3\}$, we see that \eqref{eq:1-de-plus} tends also to zero as $x\to 0$. This shows that function $N_\varphi$ belongs to the class $\rC^{q+1}$ and its partial derivatives up to order $q+1$ all vanish at zero. This proves the claim by induction. \\
Let $q$ be fixed and take a real analytic function $\varphi : \cW\dans\C$ with valuation $q$. Assume now that $N_\varphi$ belongs to the class $\rC^{q+1}$. Shrinking possibly $\cW$, we may assume that, for $x\in\cW$, 
\[\varphi (x)\ =\ \sum _{|\alpha|=q}\, a_\alpha\, x^\alpha\, +\, \tilde\varphi (x)\comma\]
where the $a_\alpha$ are complex constants and where the function $\tilde\varphi$ is either zero identically or a 
real analytic function with valuation $\geq q+1$. In particular, the function $\cW\ni x\donne |x|\tilde\varphi(x)$ does belong to the class $\rC^{q+1}$. Therefore, so does also the map 
\[g\, :\ \cW\ni x\ \donne\ |x|\, \sum _{|\alpha|=q}\, a_\alpha\, x^\alpha\period\]
Using the above study of the functions $N_\alpha :\, x\donne |x|x^\alpha$ with $\alpha\in\N^3$ and $|\alpha|=q$, we get, for all $\omega\in\SSS^2$ and $\epsilon\in\{-1; 1\}$, that $(\omega\cdot\nabla _x)^{q+1}g$ only depends on $\hat{x}=x/|x|$, that 
\[\bigl((\omega\cdot\nabla _x)^{q+1}g\bigr)_{|\hat{x}=\epsilon\omega}\ =\ \sum _{|\alpha|=q}\, a_\alpha\, \alpha !\, \bigl(|\alpha|+1\bigr)\, \omega^\alpha\, \epsilon\comma\]
and, by continuity at $0$, 
\[\sum _{|\alpha|=q}\, a_\alpha\, \alpha !\, \bigl(|\alpha|+1\bigr)\, \omega^\alpha\ =\ 0\period\]
Since the maps $\SSS^2\ni\omega\donne\omega^{\alpha}$, for $|\alpha|=q$, are linearly independent, it follows that the $a_\alpha$ are all zero. This contradicts the fact that $q$ is the valuation of $\varphi$. This proves that $N_\varphi$ does not belong to the class $\rC^{q+1}$. $\cqfd$

\Pfof{Lemma~\ref{lm:fourier-norme-localisée}}Let $\xi\in\R^3$. By definition, 
\[F_f(\xi)\ =\ \int_{\R^3}\, e^{-i\, x\cdot\xi}\, |x|\, \tau\bigl(|x|\bigr)\, dx\]
and, since $f$ is radial, $F_f$ is invariant by rotations centered at $0$. Thus $F_f(\xi)$ only depends on $\lambda :=|\xi|$. 
Let $v$ be the vector in $\R^3$ such that its coordinates in the canonical basis of $\R^3$ are $(0; 0; 1)$. In particular, $F_f(\xi)=F_f(\lambda v)$ and, changing variables into spherical coordinates, we get 
\begin{align*}
 F_f(\xi)\ =&\ \int_{\R^3}\, e^{-i\, x\cdot \lambda v}\, |x|\, \tau\bigl(|x|\bigr)\, dx\\
 \ =&\ \int_0^{+\infty}\, \tau (r)\, r^3\, \left(\int_0^{2\pi}\, d\tilde\theta\, \int_0^\pi\, e^{-ir\lambda\cos(\theta)}\, \sin (\theta)\, d\theta\right)\, dr\\
 \ =&\ 2\pi\, \int_0^{+\infty}\, \tau (r)\, r^3\, \left(\int_{-1}^1\, e^{-ir\lambda s}\, ds\right)\, dr\period
\end{align*}
Assuming $\lambda :=|\xi|>0$, we continue the computation in the following way:
\begin{align}
 F_f(\xi)\ =&\ \frac{2i\pi}{\lambda}\, \int_0^{+\infty}\, \tau (r)\, r^2\, \left(\int_{-1}^1\, (-ir\lambda)\, e^{-ir\lambda s}\, ds\right)\, dr\nonumber\\
 \ =&\ \frac{2i\pi}{\lambda}\, \int_0^{+\infty}\, \tau (r)\, r^2\, \left[e^{-ir\lambda s}\right]_{s=-1}^{s=1}\, dr\nonumber\\
 \ =&\ \frac{2i\pi}{\lambda}\, \int_0^{+\infty}\, \tau (r)\, r^2\, (-2i)\, \sin\bigl(r\lambda\bigr)\, dr\nonumber\\
 \ =&\ \frac{4\pi}{\lambda}\, \int_0^{+\infty}\, \tau (r)\, r^2\, \sin\bigl(r\lambda\bigr)\, dr\comma\label{eq:formule-fourier-norme-localisée}
\end{align}
yielding \eqref{eq:fourier-norme-localisée}. Using integration by parts, we can write 
\begin{align*}
 F_f(\xi)\ =&\ \frac{4\pi}{\lambda}\, \int_0^{+\infty}\, \tau (r)\, r^2\, \frac{1}{\lambda}\, \partial_r\, \Bigl(-\cos\bigl(r\lambda\bigr)\Bigr)\, dr\\
 \ =&\ \frac{4\pi}{\lambda ^2}\, \left(\bigl[-\tau (r)\, r^2\, \cos\bigl(r\lambda\bigr)\bigr]_{r=0}^{r=+\infty}\, +\, \int_0^{+\infty}\, \cos\bigl(r\lambda\bigr)\, \partial_r\, \bigl(\tau (r)\, r^2\bigr)\, dr\right)\\
 \ =&\ \frac{4\pi}{\lambda ^2}\, \int_0^{+\infty}\, \frac{1}{\lambda}\, \partial_r\, \Bigl(\sin\bigl(r\lambda\bigr)\Bigr)\, \partial_r\, \bigl(\tau (r)\, r^2\bigr)\, dr\\
 \ =&\ \frac{4\pi}{\lambda ^3}\, \left(\Bigl[\sin\bigl(r\lambda\bigr)\, \partial_r\, \bigl(\tau (r)\, r^2\bigr)\Bigr]_{r=0}^{r=+\infty}\, -\, \int_0^{+\infty}\, \sin\bigl(r\lambda\bigr)\, \partial_r^2\, \bigl(\tau (r)\, r^2\bigr)\, dr\right)\\
 \ =&\ \frac{-4\pi}{\lambda ^3}\, \int_0^{+\infty}\, \frac{1}{\lambda}\, \partial_r\, \Bigl(-\cos\bigl(r\lambda\bigr)\Bigr)\; \partial_r^2\, \bigl(\tau (r)\, r^2\bigr)\, dr\\
 \ =&\ \frac{-4\pi}{\lambda ^4}\, \left(\Bigl[-\cos\bigl(r\lambda\bigr)\, \partial_r^2\, \bigl(\tau (r)\, r^2\bigr)\Bigr]_{r=0}^{r=+\infty}\, +\, \int_0^{+\infty}\, \cos\bigl(r\lambda\bigr)\, \partial_r^3\, \bigl(\tau (r)\, r^2\bigr)\, dr\right)\\
 \ =&\ \frac{-4\pi}{\lambda ^4}\, \left(2\tau (0)\, +\, \int_0^{+\infty}\, \frac{1}{\lambda}\, \partial_r\, \Bigl(\sin\bigl(r\lambda\bigr)\Bigr)\, \partial_r^3\, \bigl(\tau (r)\, r^2\bigr)\, dr\right)\\
 \ =&\ \frac{-8\pi}{\lambda ^4}\, \tau (0)\, +\, \frac{-4\pi}{\lambda ^5}\, \left(0\, -\, \int_0^{+\infty}\, \sin\bigl(r\lambda\bigr)\, \partial_r^4\, \bigl(\tau (r)\, r^2\bigr)\, dr\right)\period
\end{align*}
Setting 
\[G(\xi)\ :=\ \frac{4\pi}{\lambda ^5}\, \int_0^{+\infty}\, \sin\bigl(r\lambda\bigr)\, \partial_r^4\, \bigl(\tau (r)\, r^2\bigr)\, dr\comma\]
we have $F_f(\xi)=G(\xi)-8\pi\lambda^{-4}$. By standard derivation under the integral sign, the function $G$ is smooth on $\R^3\setminus\{0\}$ and we see that the estimate in \eqref{eq:fourier-norme-localisée-symbole-2} holds true for $k=5$. \\
Integrating again by parts in the formula defining $G$ this time, we get, since $\tau$ is flat near $0$ (i.e. all the derivatives of $\tau$ vanish at $0$), that 
\begin{align}
 G(\xi)\ \ =&\ \frac{4\pi}{\lambda ^5}\, \int_0^{+\infty}\, \frac{1}{\lambda}\, \partial_r\,\Bigl(-\cos\bigl(r\lambda\bigr)\Bigr)\, \partial_r^4\, \bigl(\tau (r)\, r^2\bigr)\, dr\comma\nonumber\\
 \ =&\ \frac{4\pi}{\lambda ^6}\, \int_0^{+\infty}\, \cos\bigl(r\lambda\bigr)\, \partial_r^5\, \bigl(\tau (r)\, r^2\bigr)\, dr\label{eq:G}
\end{align}
where the integral is uniformly bounded w.r.t. $\lambda$. This yields the estimate in \eqref{eq:fourier-norme-localisée-symbole-2} for $k=6$ and $\alpha=0$. By standard derivation under the integral, we get the same estimate for $k=6$ and all $\alpha\in\N^3$. \\
Integrating by parts in a similar way in \eqref{eq:G}, we can write $G(\xi)$ as $4\pi/\lambda^k$ times some explicit integral, that is uniformly bounded w.r.t. $\lambda$. This gives the estimate in \eqref{eq:fourier-norme-localisée-symbole-2} for any $k\geq 6$ and $\alpha=0$. Standard derivation under this integral gives also this estimate for any $k\in\N$ with $k\geq 6$ and any $\alpha\in\N^3$. \\
Finally, using the explicitly known partial derivatives of the function $\R^3\setminus\{0\}\ni\xi\donne |\xi|^{-4}$, 
we observe that \eqref{eq:fourier-norme-localisée-symbole-2} implies \eqref{eq:fourier-norme-localisée-symbole-1}. 
$\cqfd$

\Pfof{Lemma~\ref{lm:manque-décroissance}} As already pointed out, we may replace the matrix $\gamma _1$ by $\gamma$. 
Let us assume that there exists some cut-off function $\tau$ and an open cone $\Gamma$ about $(-\omega ; \omega)$ such that \eqref{eq:décroiss-poly-illimitée} with $\gamma _1$ replaced by $\gamma$ holds true. We observe that the hypothesis on $\omega$ does not depend on the cut-off function $\chi _0$. In particular, we may shrink its support, preserving the requirements $\chi\chi_0=\chi_0$ and $\chi_0=1$ near $(\hat{x}; \hat{x})$. We do so to ensure that the inequality $|\tau |\geq |\tau (\hat{x}; \hat{x})|/2>0$ holds true on the support of $\tau _0 : \R^6\ni (x; x')\donne\chi _0(x)\chi _0(x')$, the cut-off function used in the definition of $F$ (cf. \eqref{eq:transf-fourier}). \\
Thanks to the previous inequality, $\tau _1 :=\tau_0/\tau$ is a well-defined, smooth, and compactly supported function, and we can write $\tau _0\gamma=\tau _1(\tau\gamma)$. The Fourier transform $F_{\tau_0\gamma}$ of $\tau_0\gamma$ is thus given by the convolution $F_{\tau_1}\ast F_{\tau\gamma}$ of the Fourier transform $F_{\tau_1}$ of $\tau_1$ with the Fourier transform $F_{\tau\gamma}$ of $\tau\gamma$. For $Y\in\R^6$, we have 
\begin{equation}\label{eq:convolution}
 F_{\tau_0\gamma}(Y)\ =\ \int_{\R^6}\, F_{\tau_1}\bigl(Y-Z\bigr)\, F_{\tau\gamma}(Z)\, dZ\period
\end{equation}
Since $\tau _1\in\rC_c^\infty(\R^6)$, we have 
\begin{equation}\label{eq:décroissance-schwartz}
\forall\, p\in\N\comma\ \sup _{Y\in\R^6}\, \bigl(1\, +\, |Y|\bigr)^p\, \bigl|F_{\tau_1}(Y)\bigr|\ <\ \infty\period
\end{equation}
We know that Fourier transform $F_{\tau\gamma}$ is bounded, that is
\begin{equation}\label{eq:paley-wiener-schwartz}
\sup _{Y\in\R^6}\ \bigl|F_{\tau\gamma}(Y)\bigr|\ <\ \infty\period
\end{equation}
We split the domain of integration in \eqref{eq:convolution} into two disjoints regions: 
\[\cR\ =\ \bigl\{Z\in\R^6\, ;\hspace{.4cm}|Y-Z|\leq \epsilon |Z|\bigr\}\hspace{.4cm}\mbox{and}\hspace{.4cm}\R^6\setminus\cR\comma\]
where $\epsilon\in ]0; 1[$. Note that, if $Z\in\cR$, then $|Y|/(1+\epsilon)\leq |Z|\leq |Y|/(1-\epsilon)$. 
We choose the parameter $\epsilon\in ]0; 1[$ so small that, for $Y\in\R(-\omega ; \omega)$ (the line generated by $(-\omega ; \omega)$), $\cR\subset\Gamma$. We observe that, there exists $\delta>0$ such that, for $|Y|>1$, for $Z\in(\R^6\setminus\cR)$, 
\begin{equation}\label{eq:taille}
 |Y-Z|\ \geq\ \delta \bigl(|Y|\, +\, |Z|\bigr)\period
\end{equation}
Let $q\in\N$, $|Y|\geq 1$, and $Y/|Y|=(-\omega ; \omega)$. There exists $D>0$ such that, for $Z\in\cR$, 
\[|Y|^q\, \bigl|F_{\tau_1}\bigl(Y-Z\bigr)\, F_{\tau\gamma}(Z)\bigr|\leq D\, \bigl(1\, +\, |Z|\bigr)^{-7}\comma\]
thanks to \eqref{eq:décroiss-poly-illimitée} and \eqref{eq:décroissance-schwartz} for $p=0$. Using \eqref{eq:taille}, \eqref{eq:paley-wiener-schwartz}, and \eqref{eq:décroissance-schwartz} for $p=q+7$, we can find $D'>0$ such that, for $Z\in(\R^6\setminus\cR)$, 
\[|Y|^q\, \bigl|F_{\tau_1}\bigl(Y-Z\bigr)\, F_{\tau\gamma}(Z)\bigr|\leq D'\, \bigl(1\, +\, |Z|\bigr)^{-7}\period\]
Using these bounds in \eqref{eq:convolution}, we get that, for all $q\in\N$, 
\[\sup _{\lambda\geq 1}\, \lambda^q\, \bigl|F_{\tau_0\gamma}\bigl(-\lambda\, \omega ; \lambda\, \omega\bigr)\bigr|\ <\ +\infty\comma\]
that is a contradiction to \eqref{eq:ordre-lambda-2} because of the choice of $\omega$. Therefore, we have $(\hat{x}; \hat{x}; -\omega; \omega)\in{\rm WF}(\gamma)$, yielding $(\hat{x}; \hat{x}; -\omega; \omega)\in{\rm WF}(\gamma _1)$. $\cqfd$

\end{appendices}
%


%
%
\end{document}